\documentclass{aa}
\usepackage{txfonts}
\usepackage{epsfig,graphics,aalongtable,amssymb}

\usepackage{natbib}
\bibpunct{(}{)}{;}{a}{}{,} 
\begin{document}

\title{Magnetic field measurements and wind-line variability\\ of OB-type stars\thanks{Figures \ref{IUEb1}-\ref{IUEo} and Table \ref{Bfieldslist} are only available in electronic form via http://www.edpsciences.org}}

\author{R.S. Schnerr\inst{1,2}
      \and H.F.~Henrichs\inst{2}
      \and C.~Neiner\inst{3}
      \and E.~Verdugo\inst{4}
      \and J.~de Jong\inst{5}
      \and V.C.~Geers\inst{2,6}
      \and K.~Wiersema\inst{7,2}
      \and B.\,van\,Dalen\inst{2}
      \and A.~Tijani\inst{2}
      \and B.~Plaggenborg\inst{2}
      \and K.L.J.~Rygl\inst{2,8}
}

\institute{SRON, Netherlands Institute for Space Research, Sorbonnelaan 2, 3584 CA Utrecht, the Netherlands
  \and     Astronomical Institute ``Anton Pannekoek'', University of Amsterdam, Kruislaan 403, 1098 SJ Amsterdam, the Netherlands
  \and     GEPI, UMR 8111 du CNRS, Observatoire de Paris-Meudon, 5 place Jules Janssen, 92195 Meudon Cedex, France
  \and     European Space Astronomy Centre (ESAC), Research \& Scientific
           Support Department of ESA, Villafranca del Castillo, P.O. Box 50727, 28080 Madrid, Spain
  \and     European Southern Observatory, Karl Schwarzschildstrasse 2, D-85748 Garching bei M\"{u}nchen
  \and     Leiden Observatory, University of Leiden, P.O. Box 9513, 2300 RA, Leiden, Netherlands
  \and     Department of Physics and Astronomy, University of Leicester, University Road, Leicester LE1 7RH, UK 
  \and     Max-Planck-Institut f\"{u}r Radioastronomie, Auf dem H\"{u}gel 69, 53121 Bonn, Germany
}

\offprints{R.S.\ Schnerr,
\email{rschnerr@science.uva.nl}}

\date{Received date / Accepted date}


\abstract
{The first magnetic fields in O- and B-type stars that do not belong to the Bp-star class, have been discovered. The cyclic UV wind-line variability, which has been observed in a significant fraction of early-type stars, is likely to be related to such magnetic fields.}
{We attempt to improve our understanding of massive-star magnetic fields, and observe twenty-five carefully-selected, OB-type stars.}
{Of these stars we obtain 136 magnetic field strength measurements. We present the UV wind-line variability of all selected targets and summarise spectropolarimetric observations acquired using the MUSICOS spectropolarimeter, mounted at the TBL, Pic du Midi, between December 1998 and November 2004. From the average Stokes $I$ and $V$ line profiles, derived using the LSD method, we measure the magnetic field strengths, radial velocities, and first moment of the line profiles.}
{No significant magnetic field is detected in any OB-type star that we observed. Typical 1$\sigma$ errors are between 15 and 200 G. A possible magnetic-field detection for the O9V star 10 Lac remains uncertain, because the field measurements depend critically on the fringe-effect correction in the Stokes $V$ spectra.
We find excess emission in UV-wind lines, centred about the rest wavelength, to be a new indirect indicator of the presence of a magnetic field in early B-type stars. The most promising candidates to host magnetic fields are the B-type stars $\delta$ Cet and 6 Cep, and a number of O stars.}
{Although some O and B stars have strong dipolar field, which cause periodic variability in the UV wind-lines, such strong fields are not widespread. If the variability observed in the UV wind-lines of OB stars is generally caused by surface magnetic fields, these fields are either weak ($\lesssim$ few hundred G) or localised.}

\authorrunning{R.S. Schnerr et al.}
\maketitle



\section{Introduction}

Magnetic fields play an important role in many astrophysical contexts. They
have been discovered at all stages of stellar evolution. Fields of the order of
$\mu$G to mG have been measured in star-forming molecular clouds, which are
dynamically important during cloud collapse \citep[e.g.][]{crutcher:1999}. The young T Tauri stars have magnetic fields
that guide accreting matter into the inner part of disk
\citep[e.g.][]{valenti:2004}, and the first magnetic field detections for the accreting Herbig Ae/Be stars have also been
reported \citep[][]{hubrig:2004,wade:2005,hubrig:2006b,catala:2007}, following a marginal detection for HD 104237 \citep{donati:1997}. For main sequence stars, magnetic
fields have been found in late-type stars, which are thought to have
dynamo-generated fields, and early-type stars, such as the strongly magnetic
Ap/Bp stars \citep[see][for an overview]{mathys:2001}. The end products of
stellar evolution, white dwarfs and neutron stars, have been found to have
very strong ($10^6-10^{15}$G) magnetic fields \citep[see][for a review]{wickramasinghe:2000,manchester:2004}. It is not known whether all
new-born neutron stars are strongly magnetic, but certainly a very significant
fraction apparently is. The immediate (unsolved) question arises how these
neutron stars obtained their magnetic field: did their progenitors (the O and B
stars) have no significant field and is the field generated just after the
collapse, or did they possess a field when they were born, which survived
during their life, and which is then strongly amplified during the core collapse? In the massive OB stars ($>$9 M$_{\odot}$)
fields are not generated by contemporary dynamos such as in low-mass
main-sequence stars. Fossil fields, which originate in the interstellar medium, could
survive the radiative phase during contraction, because these stars
do not become fully convective. This was indeed proposed by \citet{ferrario:2005}, who in addition argue that conservation of a significant fraction of the magnetic
flux of the massive stars during their lives is consistent with the strong
fields observed in neutron stars, analogous to the origin of the
magnetic fields of the strongly magnetic white dwarfs. Constraints on magnetic
fields in rotating massive stars with winds were discussed by
\citet{maheswaran:1992}. Mechanisms that can generate a field during the main-sequence phase either in the convective core
\citep{charbonneau:2001} or in shear-unstable radiative layers \citep{macdonald:2004, mullan:2005} were investigated. The long-term effects of magnetic fields on the stellar interior were studied by, e.g., \citet{spruit:2002} and
\citet{maeder:2003,maeder:2004}. The work by \citet{heger:2005} demonstrated the
dramatic influence of incorporating magnetic fields into stellar evolution models prior to core collapse. Simple magnetic-flux conservation arguments indicate that the observed field strengths in neutron stars of 10$^{12}$ G can be attained easily by a progenitor star that has a surface magnetic-field strength of 100 G or even less.
The main difficulty with this scenario, however, is that such fields have never been
detected, the most likely reason being that the expected strength is below the
detection limit of most current instruments.

\begin{table*}[ht!]
\caption{Known magnetic massive OB stars and their properties. We note that for a dipole field, the magnetic field strength at the pole ($B_{\rm pol}$) and the equator ($B_{\rm eq}$) are related by $B_{\rm pol}=2\,B_{\rm eq}$. }
\label{magnetic}
\begin{center}
\begin{tabular}{llrrrrlll}
\hline
\hline
Name & Spectral Type & $v\sin i$ & \multicolumn{1}{c}{$P_{\rm rot}$} & \multicolumn{1}{c}{$M$}   &
\multicolumn{1}{c}{Inclination} & \multicolumn{1}{c}{$\beta$} & \multicolumn{1}{c}{$B_{\rm pol}$} & Reference\\
 &  & \multicolumn{1}{c}{(km s$^{-1}$)}  &  \multicolumn{1}{c}{(d)}  & \multicolumn{1}{c}{($M_{\odot}$)}  & \multicolumn{1}{c}{(deg.)} & \multicolumn{1}{c}{(deg.)}   &  \multicolumn{1}{c}{(gauss)} &\\
\hline
$\theta^1$Ori C & O4-6V  & 20             & 15.4  &	  45 & $\sim$45  & 42$\pm$6  & 1100$\pm$100  & \citet{donati:2002}\\
HD 191612       & Of?p   & $<$77          & 538$^{\rm a}$ & $\sim$40 & $\sim$45  & $\sim$45  &  $\sim$1500   & \citet{donati:2006}\\
$\tau$ Sco      & B0.2V  & 5$^{\rm b}$    & 41    &	  15 & $\sim$70  & $\sim$90  & $\sim$500$^{\rm c}$ & \citet{donati:2006b}\\
$\xi^1$ CMa	& B0.5IV & 20     	  & $<$37 &	  14 &  	 &	     & $\sim$500     & \citet{hubrig:2006}\\
$\beta$ Cep	& B1IV   & 27     	  & 12.00 &	  12 & 60$\pm$10 & 85$\pm$10 & 360$\pm$40    & \citet{henrichs:2000a}\\
V2052 Oph       & B1V    & 63     	  & 3.64  &	  10 & 71$\pm$10 & 35$\pm$17 & 250$\pm$190   & \citet{coralie:2003c}\\
$\zeta$ Cas	& B2IV   & 17     	  & 5.37  &	   9 & 18$\pm$4  & 80$\pm$4  & 340$\pm$90    & \citet{coralie:2003a}\\
$\omega$ Ori    & B2IVe  & 172    	  & 1.29  &	   8 & 42$\pm$7  & 50$\pm$25 & 530$\pm$200   & \citet{coralie:2003b}\\
\hline
\hline
\multicolumn{9}{l}{$^{\mathrm{a}}$ To be confirmed}\\
\multicolumn{9}{l}{$^{\mathrm{b}}$ \citet{mokiem:2005}}\\
\multicolumn{9}{l}{$^{\mathrm{c}}$ Field more complex than dipolar}\\
\end{tabular}
\end{center}
\end{table*}

\citet{braithwaite:2004b} showed that the kG fields found in the Ap/Bp stars are probably fossil remnants of star formation. This is supported by the discovery of magnetic fields in Herbig Ae/Be stars, as discussed above.
If this result can be extrapolated to more massive B and O stars, which also have stable, radiative envelopes, one would expect to find more magnetic, massive stars than the few examples discovered so far (see Table \ref{magnetic}). Considering the widespread phenomena observed in massive stars that can attributed to magnetic fields, such as UV wind-line variability, unusual X-ray spectra, H$\alpha$ variability and non-thermal radio emission \citep[as summarised by][]{henrichs:2005}, a comprehensive study to confirm the presence of magnetic fields is justified.

To obtain more insight into the fraction of magnetic OB-type stars and the strengths of their magnetic fields, we selected a group of OB stars that show indirect indications of a magnetic field. In Sect.~\ref{indicators}, we discuss the indirect indicators for all program stars, in particular the stellar wind variability and abundances. In Sect.~\ref{obs&data}, we describe how we obtained circular polarisation spectra which enable the determination of the longitudinal component of the magnetic field integrated over the stellar disk, and discuss the observations and data reduction procedure. In Sect.~\ref{sect:results} and \ref{conclusions}, we present the results and conclusions that can be drawn from this survey.

\begin{table*}[ht!]
\caption[Targets.]{Properties of program stars in this survey with the integration limits used to determine the magnetic-field strength. Rotational velocities are taken from the Bright Star Catalogue \citep[][O stars]{bsc5:1991}, \citet[][B stars]{abt:2002} and \citet[][A stars]{abt:1995}, unless indicated otherwise. Spectral types are from \citet{walborn:1972,walborn:1973} and \citet{walborn:1990}. Selection criteria are UV-line variability (UV), nitrogen abundance anomalies \citep[N,][]{gies:1992}, detection in X-rays by ROSAT \citep[][X]{berghofer:1996}, and known $\beta$ Cep-type pulsator (var).}
\label{targetlist}
\begin{center}
\begin{tabular}{rllrrrrcc}
\hline
\hline
\multicolumn{1}{c}{HD} & Star & Spectral    & \multicolumn{1}{c}{Nr.}  & \multicolumn{1}{c}{$v\sin i$} & \multicolumn{1}{c}{$v_{\rm rad}$}& Integration limits               &Years & Selection\\
number                 &      & Type        & \multicolumn{1}{c}{data sets} & (km s$^{-1}$)            & \multicolumn{1}{c}{(km s$^{-1}$)} & \multicolumn{1}{c}{(km s$^{-1}$)} &      & criteria\\
\hline
\multicolumn{9}{c}{\bf{B stars}}\\
   886 & \object{$\gamma$ Peg}         & B2IV         & 2 &   0     & 4.1     & [$-$37.4,+37.4] &2002           & var\\
 16582 & \object{$\delta$ Cet}         & B2IV         & 1 &   5     & 13.0    & [$-$47.4,+47.4] &2003           & N,var\\
 37042 & \object{$\theta^2$ Ori B$^a$} & B0.5V        & 1 &  50$^b$ & 28.5    & [$-$77,+77]     &2004           & X\\
 74280 & \object{$\eta$ Hya}           & B3V          & 3 &  95$^*$ & 21      & [$-$192,+196]   &1998           & var\\
 87901 & \object{$\alpha$ Leo}         & B7V          & 1 & 300$^*$ & 5.9     & [$-$452,+452]   &1998           & cal\\
 89688 & \object{RS Sex}               & B2.5IV       & 2 & 215     & 5       & [$-$350,+350]	&1998		& var\\
116658 & \object{$\alpha$ Vir}         & B1III-IV+B2V & 1 & 130     & 1       & [$-$368,+264]	&2000		& X,var\\
144206 & \object{$\upsilon$ Her }      & B9III        & 3 &  20     & 2.7     & [$-$21,+21]	&2001		& abun\\
147394 & \object{$\tau$ Her}           & B5IV         &35 &  30$^*$ & $-$13.8 & [$-$107,+105]   &2001/2002/2003 & abun\\
160762 & \object{$\iota$ Her}          & B3IV         & 2 &   0     & $-$20.0 & [$-$18.4,+19.0] &2001		& UV,var\\
182568 & \object{2 Cyg}                & B3IV         & 1 & 100     & $-$21   & [$-$244,+203]	&2003		& abun\\
199140 & \object{BW Vul}               & B2IIIe       & 5 &  45     & $-$6.1  & [$-$159,+83]	&2002		& UV,var\\
203467 & \object{6 Cep}                & B3IVe        & 1 & 120     & $-$18   & [$-$222,+249]	&2002		& UV\\
207330 & \object{$\pi^2$ Cyg}          & B3III        &15 &  30     & $-$12.3 & [$-$96.6,+96.6] &2001/2002	& N\\
217675 & \object{o And}                & B6IIIpe+A2p  & 1 & 200     & $-$14.0 & [$-$430,+500]	&2002		& UV\\
218376 & \object{1 Cas}                & B0.5IV       &18 &  15     & $-$8.5  & [$-$94,+94]	&2001/2002	& UV,N\\
\multicolumn{9}{c}{\bf{B supergiants}}\\
 34085 & \object{$\beta$ Ori}          & B8Ia         & 4 &  40     & 20.7    & [$-$87.8,+87.8] &2004		& UV,X\\
 91316 & \object{$\rho$ Leo}           & B1Iab        & 2 &  50     & 42.0    & [$-$116,+116]	&1998		& UV\\
164353 & \object{67 Oph}               & B5Ib         & 6 &  40     & $-$4.7  & [$-$75,+86]	&2002		& UV\\
\multicolumn{9}{c}{\bf{O stars}}\\
 30614 & \object{$\alpha$ Cam}         & O9.5Ia       & 4 &  95     & 6.1     & [$-$203,+203]	&1998		& UV,X\\
 34078 & \object{AE Aur}               & O9.5V        & 1 &   5     & 59.1    & [$-$70,+70]	&1998		& UV,X\\
 36861 & \object{$\lambda$ Ori A}      & O8III((f))   & 4 &  66     & 33.5    & [$-$170,+170]	&2004		& UV\\
 47839 & \object{15 Mon}               & O7V((f))     & 5 &  63     & 33.2    & [$-$168,+168]	&1998/2004	& UV,X\\
149757 & \object{$\zeta$ Oph}          & O9.5Vnn$^c$  & 3 & 379     & $-$15   & [$-$664,+664]	&2001/2002	& UV,X\\
214680 & \object{10 Lac}               & O9V          &15 &  31     & $-$9.7  & [$-$83,+83]	&1998/2003/2004 & UV,X\\
\multicolumn{9}{c}{\bf{Magnetic calibration stars and other targets}}\\
 65339 & \object{53 Cam}               & A2pSrCrEu    & 1 &  15       & $-$4.8  & [$-$51,+60]     &1998           & \\
112413 & \object{$\alpha^{2}$ CVn}     & A0pSiEuHg    & 8 & $<$10     & $-$3.3  & [$-$31,+44]     &2000/2001/2003 & \\
182989 & \object{RR Lyrae}             & F5           & 9 & $<$10$^d$ & $-$72.4 & [$-$48,+33]     &2003           & \\
\hline
\hline
\end{tabular}
\end{center}
\begin{list}{}{}
\item[$^*$] Rotational standard of \citet{slettebak:1975}
\item[$^{\mathrm{a}}$] \citet{houk:1999}
\item[$^{\mathrm{b}}$] \citet{wolff:2004}
\item[$^{\mathrm{c}}$] \citet{maiz:2004}
\item[$^{\mathrm{d}}$] \citet{peterson:1996}
\end{list}
\end{table*}

\section{Indirect magnetic-field indicators and target selection}
\label{indicators}
\citet{cassinelli:1985} presented the first survey of evidence that magnetic fields in the atmospheres of massive stars were the most probable explanation of observed non-radiative activity. \citet{henrichs:2005} discussed a number of unexplained observational phenomena in massive stars that could be considered as indirect indicators of the presence of a stellar magnetic field. We use the indicators as criteria to select our targets in this study. We include primarily targets selected on the basis of their UV wind-line variability, abundance anomalies and X-ray emission, and some $\beta$ Cephei pulsators, as indicated in the last column of Table \ref{targetlist}. A further selection was made according to the location of the observatory, observing season, and a favourable position in the sky at the time when no other higher priority targets were observable. We discuss the background behind each of these selection criteria.

\subsection{Indirect indicators}

\subsubsection{Stellar wind variability}
\label{windvariability}

Wind variability has proven to be a particularly successful indirect indicator, as demonstrated by the discovery of the magnetic OB stars $\beta$ Cep \citep{henrichs:2000a}, $\zeta$ Cas \citep{coralie:2003a}, V2052 Oph \citep{coralie:2003c}, and $\theta^1$ Ori C \citep{donati:2002}. These stars were selected because of the striking time behaviour of the UV stellar-wind lines of \ion{C}{iv}, \ion{Si}{iv} and \ion{N}{v}, which in the first three cases was characterised by a regular modulation of the entire profile, centred on the rest wavelength of the transitions, and similar to what is observed in magnetic Bp stars.

Time-resolved observations, acquired primarily using the IUE and FUSE satellites, showed that in the observed cases at least 60\% of the O stars, 17\% of the non-chemically peculiar B stars, and all of the Bp stars have variable wind-lines \citep{henrichs:2005}. For a comprehensive review of the cyclical wind-variability of O stars, see \citet{fullerton:2003}, who summarised the properties of about 25 O stars which had sufficient time-series data available. Two different categories can be distinguished. Firstly, for stars with large scale, dipole-like magnetic fields (the Bp stars and the stars in Table \ref{magnetic}), this variability is probably due to material that is guided by the magnetic field that co-rotates with the star \citep{shore:1987, schnerr:2007}. In these oblique rotators, the timescale of the variability coincides with the rotation period. Secondly, cyclical variability, with a timescale comparable to the estimated rotation period of the star, is commonly observed \citep[as summarised by][]{fullerton:2003}. In such cases the period does not keep phase over much longer periods. This is presumably the case for the majority of the early-type stars. The variability is mostly observed in the form of the Discrete Absorption Components (DACs), which are distinct absorption features that progress repeatedly bluewards through the absorption part of the P-Cygni wind profiles on a timescale of a few days, i.e.\ similar to the rotation timescale of the star. For many stars, only snapshots of UV-wind lines are available rather than timeseries; however, from the characteristic shape of the DACs, one may conclude that these stars are likely to behave in a similar way, even if the timescale is unknown. Using hydrodynamical simulations \citet{cranmer:1996} showed that DACs can be generated by magnetic footpoints on the stellar surface; this is a strong motivation for the search presented in this paper, although other azimuthal perturbations of the wind, such as non-radial pulsations, could cause similar effects. Non-radial pulsations of O stars have timescales much shorter than the DAC recurrence timescales \citep{dejong:1999,henrichs:1999}; although they could contribute, they are, for this reason, unlikely to be the main cause. \citet{kaper:1997} presented observational arguments for a magnetic origin of DACs in OB stars by studying wind and H$\alpha$ variability simultaneously.  H$\alpha$ emission, which originates close to the stellar surface, often shows covariability with the DACs; there are, for example, the well-documented cases of the O stars $\xi$~Per \citep{dejong:2001} and $\zeta$ Pup \citep{reid:1996}. A systematic search for cyclical variability, in H$\alpha$ profiles of 22 OB supergiants, was completed by \citet{morel:2004}. The general conclusion is that the DACs and H$\alpha$ variability are indicators of the same phenomenon.

For Be stars, in addition to the magnetic and DAC-type UV resonance line-variability described above, there is a third intermediate type: the variable absorptions in such cases occur at a much lower velocity than for the DACs, but are, unlike magnetic oblique rotators, found at velocities that are significantly above zero. This was shown by \citet[see also \citealt{evan:2004}]{henrichs:2005}, who concluded, after an exhaustive study of all available spectra for 81 Be stars in the IUE archive, that 57 stars exhibit no wind variability, 5 stars are of the magnetic type, 7 stars show DAC variability, and 12 belong to the intermediate type. The working hypothesis is that stars with these last three types of variability all have surface magnetic fields, but differ in both the geometry and magnetic confinement parameter $\eta$, which is the ratio of the magnetic to the wind pressure as defined by \citet{asif:2002}

\begin{equation}
    \eta \equiv { B_{\rm eq}^2/8 \pi \over \rho v^{2} /2 } \approx {B_{\rm eq}^2 R_{*}^2 \over \dot{M} v_{\infty}^{2} },
\label{etadef}
\end{equation}
in which $B_{\rm eq}$ is the equatorial field strength at the surface of the star, which has a radius $R_{*}$, mass-loss rate $\dot{M}$, and terminal wind velocity $v_{\infty}$, with wind density $\rho$.
If $\eta > 1$, the magnetic field will dominate the wind behaviour.  This would be the case for typical wind parameters in early-type stars with field strengths of more than 50\,--\,100 G. The stars listed in Table~\ref{magnetic} have $\eta$ values of about 10--100.

\begin{figure*}[tbh!]
\begin{center}
\includegraphics[width=0.9\linewidth,angle=0,trim=0 0 0 0, clip]{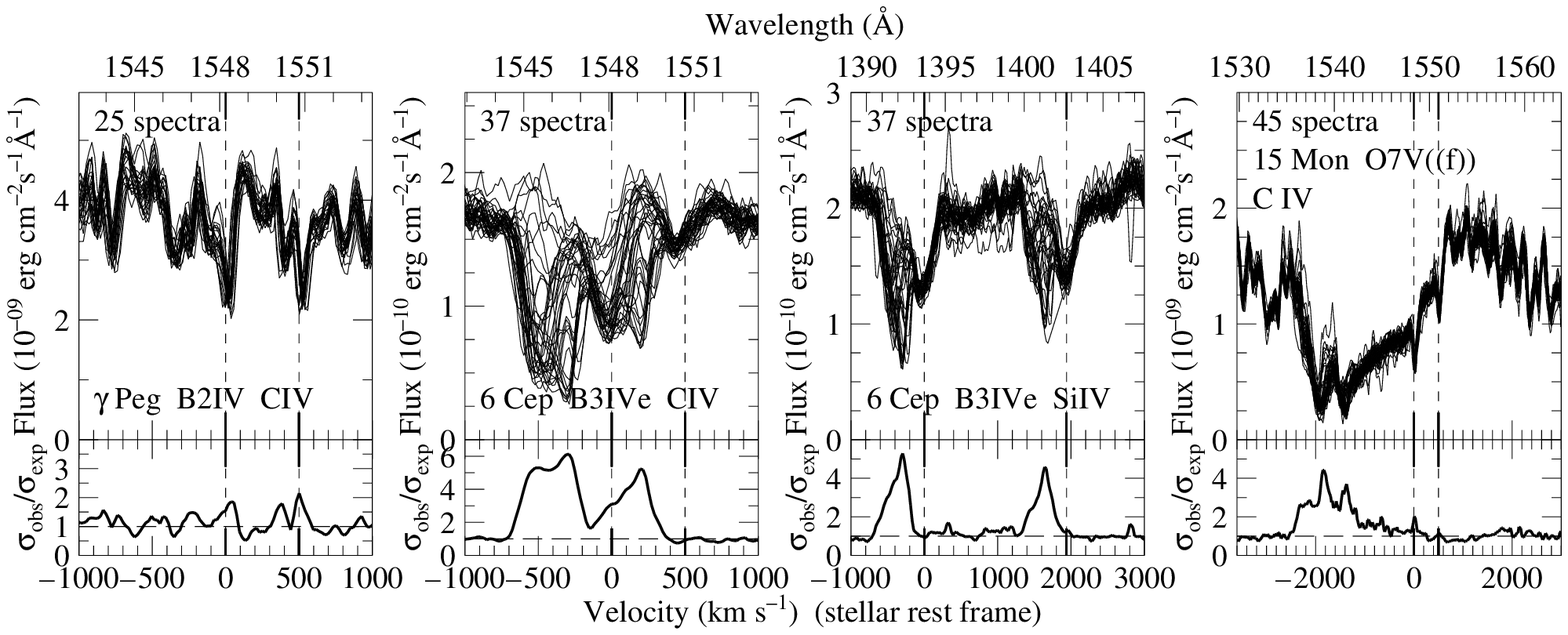}
\caption[]{Example plots of the UV variability for three stars of our sample. Plots for all stars of our sample for which IUE observations are available can be found in Figs.~\ref{IUEb1}-\ref{IUEo} of the online edition. {\bf Top panels}: UV spectra near the \ion{C}{iv} and \ion{S}{iv} resonance lines as observed with the IUE satellite. {\bf Bottom panels}: ratio of observed to the expected variance, which is a measure for the significance of the variability. The horizontal velocity scales are with respect of the rest wavelength of the principal member of the \ion{C}{iv} doublet, corrected for the radial velocity of the star as listed in Table \ref{targetlist}. Vertical dashed lines denote the positions of the rest wavelengths of the doublet members.}
\label{IUE_example}
\end{center}
\end{figure*}

\begin{figure}[htbp]
\centering
\includegraphics[width=0.95\linewidth,angle=0,trim=0 0 0 0, clip]{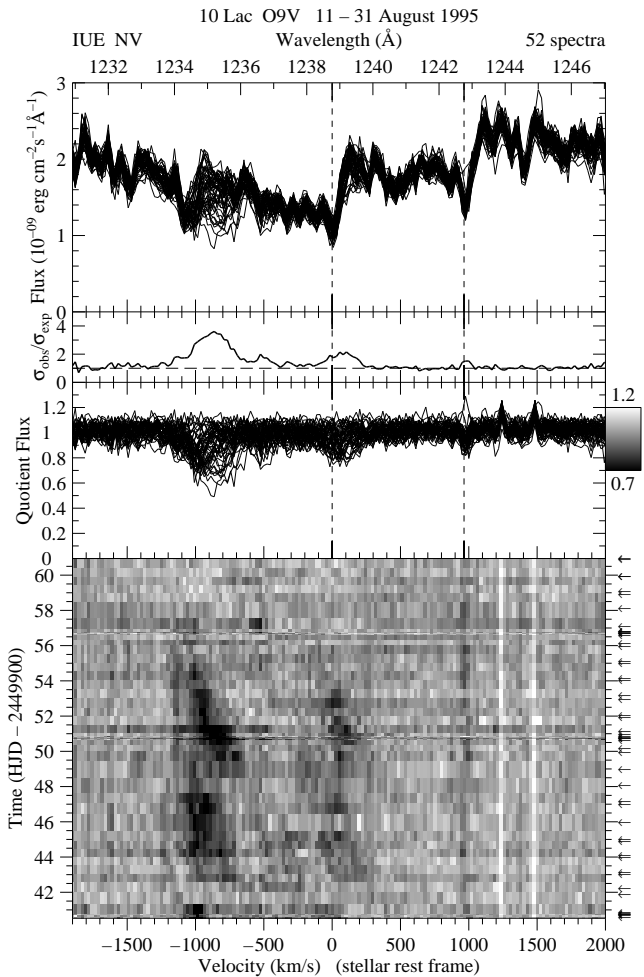}
\caption[]{Timeseries of 20 days in August 1995 of the \ion{N}{v} UV resonance lines of the slowly-rotating O9V star 10 Lac, showing the appearance and development of the Discrete Absorption Components in both doublet members. The horizontal scale and the two panels at the top are similar to Fig.~\ref{IUE_example}.
{\sl Third panel:} Overplot of quotient spectra (see text); {\sl Bottom panel:} Gray-scale representation of the quotient spectra. Arrows indicate the mid epochs of the observations.}
\label{IUE10lac}
\end{figure}

In Figs.~\ref{IUEb1}, \ref{IUEb2}, and \ref{IUEo}, we show selected \ion{C}{iv} and \ion{Si}{iv} profiles, with a measure of their variability, of all our targets for which high-resolution spectra, in the short-wavelength range, are available in the IUE archive; also if the target was not selected for this study based on its wind-line variability. As these figures are only available in the online edition, example plots for three selected stars are shown in Fig.~\ref{IUE_example}. For some stars, we show both spectral regions if they are of particular interest. Before calculating the variance, we normalised the flux values to their average value at selected portions of the continuum that were unaffected by the stellar wind, and applied the resulting scaling factor to the entire spectrum. This is required because the UV flux may vary, of which BW Vul is an extreme example, and mixed absolute-flux calibrations of images acquired using both the large and small apertures during over 18 years of operations of the IUE satellite, were sometimes left with some systematic error. Typical signal-to-noise ratios are about 20, which should be kept in mind when no variability is reported. The temporal variance spectra in the bottom panels indicate the significance of the variability \citep{henrichs:1994, fullerton:1996}. For each set of observations, a separate noise model was applied, which was adapted to the quality of the set \citep[see][]{henrichs:1994}. In Fig.~\ref{IUE10lac}, we show the development of the Discrete Absorption Components in the O9V star 10 Lac, as an example of (presumably) cyclic variability, although the duration of the observations is insufficient for the detection of recurrent behaviour. For producing the grey-scale image, we constructed quotient spectra by using a template spectrum, which was produced using the highest points of all spectra, taking noise into account. This method was developed by \citet{kaper:1999}.
Similar DAC behaviour has been observed in the magnetic O star $\theta^1$~Ori~C, where the origin of the DACs could be traced back to the north magnetic pole \citep{henrichs:2005}; this provides strong support to our hypothesis that this type of wind variability has a magnetic origin.

\begin{figure}[th!]
\centering
\includegraphics[width=0.95\linewidth,angle=0,trim=0 0 0 0, clip]{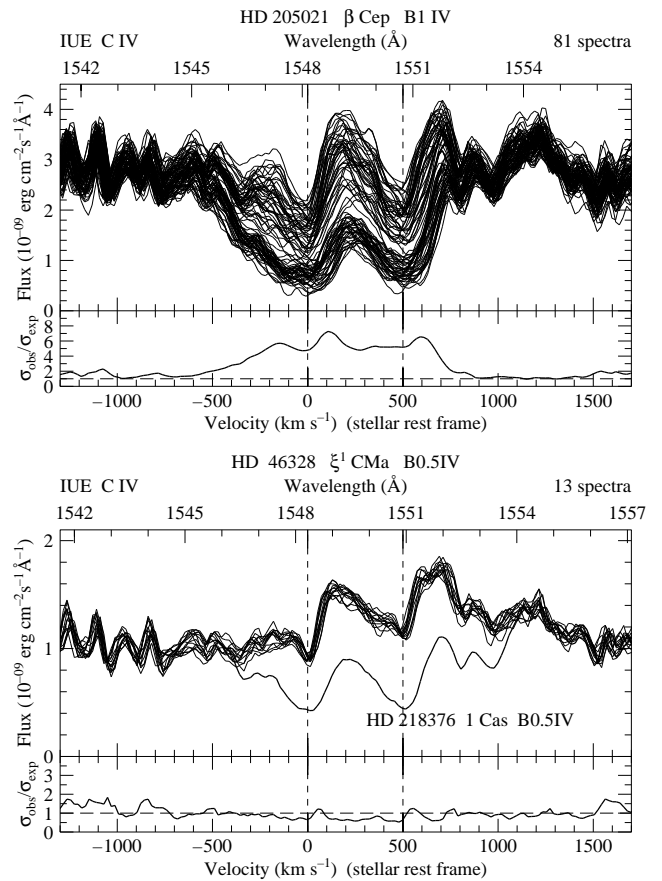}
\caption[]{Comparison between \ion{C}{iv} profiles of the magnetic oblique rotators $\beta$ Cep ({\sl top}) and $\xi^1$ CMa ({\sl bottom}). The panels are similar to Fig.~\ref{IUE_example}. To appreciate the excess emission in $\xi^1$ CMa, we overplot the scaled profile of the B0.5IV star 1 Cas, which has a typical \ion{C}{iv} profile for this spectral type.}
\label{bcepxi1cma}
\end{figure}

We note that no long UV wind-profile timeseries are available for the recently discovered magnetic B0.5IV star
$\xi^1$~CMa \citep{hubrig:2006}: apart from one isolated spectrum, the 12 remaining spectra were taken within 7 hours, and no typical modulation is present; the \ion{C}{iv} profiles are strikingly similar, however, to the profiles of $\beta$ Cep during its maximum emission phase (see Fig. \ref{bcepxi1cma} for a comparison). For $\beta$ Cep, sufficient data are available to cover several rotational periods, which show the gradual transition from an enhanced to a reduced contribution of emission centred close to zero velocity, typical for magnetic B stars. For $\xi^1$ CMa, the data span only several hours and no rotational modulation is expected to be observed, but the unusual emission profiles are similar to the most extreme emission profiles of $\beta$ Cep. In this figure (lower part), we overplotted an averaged and scaled \ion{C}{iv} profile of 1 Cas, which is a star of the same spectral type and with a comparable value of $v \sin i$, to enhance the contrast with an average B0.5IV star. Such unusual emission, centred about zero velocity, and in this case extending to about 500 km s$^{-1}$ either side, is found only for magnetic B stars, and is an additional indirect indicator if timeseries are not available.

\subsubsection{Nitrogen enhancements}
$\beta$ Cep was suspected to posses a magnetic-field because of the stars unusual wind modulation \citep{henrichs:2000a}. When its magnetic field was in fact discovered, we identified this star as one of the N-enhanced stars previously studied by \citet{gies:1992}, and decided to include other stars from their study in our observing program with the TBL at the Pic du Midi, i.e.\ \object{$\zeta$ Cas}, \object{$\delta$ Cet}, \object{$\pi^2$ Cyg} and \object{1 Cas}.
As an immediate result the star $\zeta$ Cas was found to have a magnetic field \citep{coralie:2003a}; this star demonstrated the same type of wind modulation as $\beta$ Cep. $\xi^1$ CMa, which is not in a favourable position to be observed from the TBL, shows similar variability and was found to have a magnetic field by \citet{hubrig:2006}.
\citet{morel:2006} determined abundances of several elements including nitrogen, for a number of slowly rotating $\beta$ Cephei stars. These included, in particular, $\gamma$ Peg,  \object{$\nu$ Eri}, $\delta$ Cet, \object{$\xi^1$ CMa}, V2052 Oph and $\beta$ Cep, the final four of which have N-enhanced abundances and are magnetic, oblique rotators (apart from $\delta$ Cet, see below), which is evidence of a strong correlation between magnetic fields and enhanced nitrogen abundances. Possible theoretical explanations are also discussed in this paper. We note that the B2 III star $\nu$ Eri was not found to be magnetic by \citet{schnerr:2006a}, in agreement with this correlation.
Chemical peculiarities in early B stars are often considered as an indication of a magnetic field, since these peculiarities can arise from the inhibition of mixing motion by a field. In contrast to the hotter O stars where chemical enhancements are an indicator of their evolutionary stage.

\subsubsection{X-ray properties}
Strong X-ray emission from hot stars was discovered by the Einstein mission \citep{harnden:1979, seward:1979}. Some OB stars show variable, hard X-ray emission that cannot be explained by instability-driven wind shocks; magnetospheres may then play a key role in the X-ray emission process, as shown for $\beta$ Cep by \citet{donati:2001}. \citet{schulz:2000} showed that the magnetic star $\theta^1$ Ori C has broadened X-ray line profiles, which are symmetric about their rest wavelengths, as opposed to other types of X-ray line profiles, which are narrow or blue-shifted \citep[e.g.][]{cohen:2003}. The X-ray line profiles of non-magnetic hot stars were successfully modelled by \citet{oskinova:2004,oskinova:2006} by taking the effects of clumping into account. Our target selection was based on the detection of excess X-ray emission in the ROSAT All Sky Survey \citep{berghofer:1996}. These stars are marked with an ``X'' in Table~\ref{targetlist}.

\subsection{Target selection}
\label{targetselection}

Specific remarks pertinent to most of our targets, regarding selection criteria, are listed here, in the order in which they appear in Table~\ref{targetlist}. We aimed to include the most recent references, many of which were unavailable at the time of observations; these data however often strengthen our argument to include a given star in future searches.

\subsubsection{B stars}
\begin{figure*}[tbhp]
\begin{center}
\includegraphics[width=0.9\linewidth,angle=0,trim=0 0 0 0, clip]{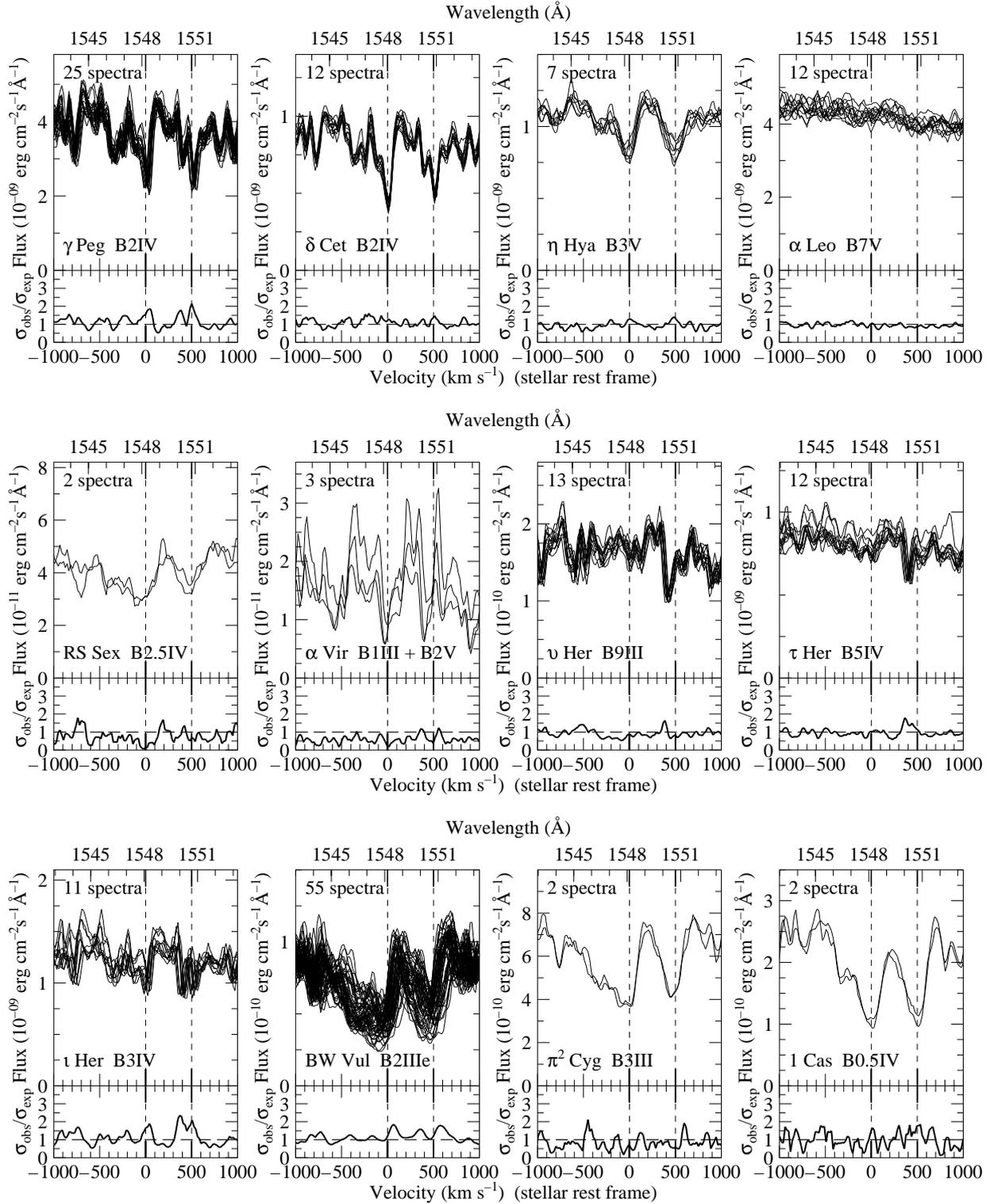}
\caption[]{Top panels: UV spectra near the \ion{C}{iv} resonance lines as observed with the IUE satellite of the first part of the B stars of our sample. Bottom panels: ratio of observed to the expected variance, which is a measure for the significance of the variability. The horizontal velocity scales are with respect of the rest wavelength of the principal member of the \ion{C}{iv} doublet, corrected for the radial velocity of the star as listed in Table \ref{targetlist}. Vertical dashed lines denote the positions of the rest wavelengths of the doublet members.}
\label{IUEb1}
\end{center}
\end{figure*}

{\sl HD 886 (\object{$\gamma$ Peg})} B2IV. A well-known $\beta$ Cep variable. The UV spectra in Fig. \ref{IUEb1} are not corrected for radial velocity due to the pulsations, and no other significant variability has been observed. \citet{peters:1976}, \citet{pintado:1993}, and \citet{gies:1992} measured approximately solar abundances of the CNO elements.

{\sl HD 16582 (\object{$\delta$ Cet})} B2IV. \citet{gies:1992} found a nitrogen excess in this multi-period $\beta$ Cephei star \citep{aerts:2006}, which was confirmed by \citet{morel:2006}. This N enhancement is similar to the
three other known magnetic $\beta$ Cephei stars, which makes this star a strong magnetic candidate, as also noted by \citet{hubrig:2006}. 

Variability similar to that typically observed for oblique rotators, is not observed in our 12 \ion{C}{iv} profiles. This is however unsurprising because all spectra were acquired within six hours; this corresponds to less than two pulsation cycles, and is significantly less then the estimated rotation period of 2 or 4 weeks, which  was measured for a pole-on viewed, slow rotator with $v\sin i \simeq$1 km~s$^{-1}$ \citep{aerts:2006}.

{\sl HD 37042 (\object{$\theta^2$ Ori B})} B0.5V. Following discussion with M.\ Gagn\'e, this target was introduced to our study based on its Chandra X-ray observations, which showed it to be a bright X-ray source.

{\sl HD 74280 (\object{$\eta$ Hya})} B3V. This $\beta$ Cep variable has a slight underabundance of carbon \citep{kodaira:1970}, but no value for the N abundance was determined.  The 7 IUE spectra are snapshots sampled over 12 years, but no significant wind-profile changes are apparent.

{\sl HD 87901 (\object{$\alpha$ Leo})} B7V. The spectra of this star were used for the correction of the fringes in the spectra. For completeness, we show the wind profiles, which do not change over 16 years.

{\sl HD 89688 (\object{RS Sex})} B2.5V. This is an unusually rapidly rotating $\beta$ Cephei star. The two UV spectra were taken 4 years apart, but show no variation.

{\sl HD 116658 (\object{$\alpha$ Vir})} B1III-IV+B2V. This X-ray emitter and $\beta $ Cephei variable is in a 4-day binary orbit. The radial-velocity shifts, rather than wind variations, are responsible for the variability in the \ion{C}{iv} line. This line was monitored over 16 years; twelve spectra were acquired during a single pulsation cycle.

{\sl HD 144206 (\object{$\upsilon$ Her})} B9III is a slowly-rotating HgMn star \citep{adelman:1992,adelman:2006}, with no obvious UV profile variations over 12 years.

{\sl HD 147394 (\object{$\tau$ Her})} B5IV. This is a slowly-pulsating B star \citep{briquet:2003} with
inconsistent measurements of its metallicity \citep[see][]{niemczura:2003,rodriguez:2005}. The UV \ion{C}{iv} profiles show no variability when the low quality of some of the earlier data are taken into account.

{\sl HD 160762 (\object{$\iota$ Her})} B3IV. This is a $\beta$ Cephei star for which \citet{kodaira:1970} reported a slight N enhancement, although \citet{pintado:1993} found solar abundances and \citet{grigsby:1996} found a significant underabundance in iron relative to the Sun. The small apparent UV profile changes in total flux are caused by calibration uncertainties of the IUE observations, which were taken using the small aperture.

{\sl HD 182568 (\object{2 Cyg})} B3IV. A He-weak star \citep{lyubimkov:2004} for which \citet{bychkov:2003} report a magnetic field measurement of $19\pm298$ G, using Balmer-line wing measurements. There is only one reliable high-resolution IUE spectrum (not included in the figures) that shows normal wind profiles.

{\sl HD 199140 (\object{BW Vul})} B2IIIe. This well-known $\beta$ Cephei star has large UV flux variations, and the normalised \ion{C}{iv} profiles show only radial-velocity shifts due to pulsation. \citet{stankov:2003} reported subsolar values for the abundance of He and some other elements, but normal values for N.

\begin{figure*}[thbp]
\begin{center}
\includegraphics[width=0.9\linewidth,angle=0,trim=0 0 0 0, clip]{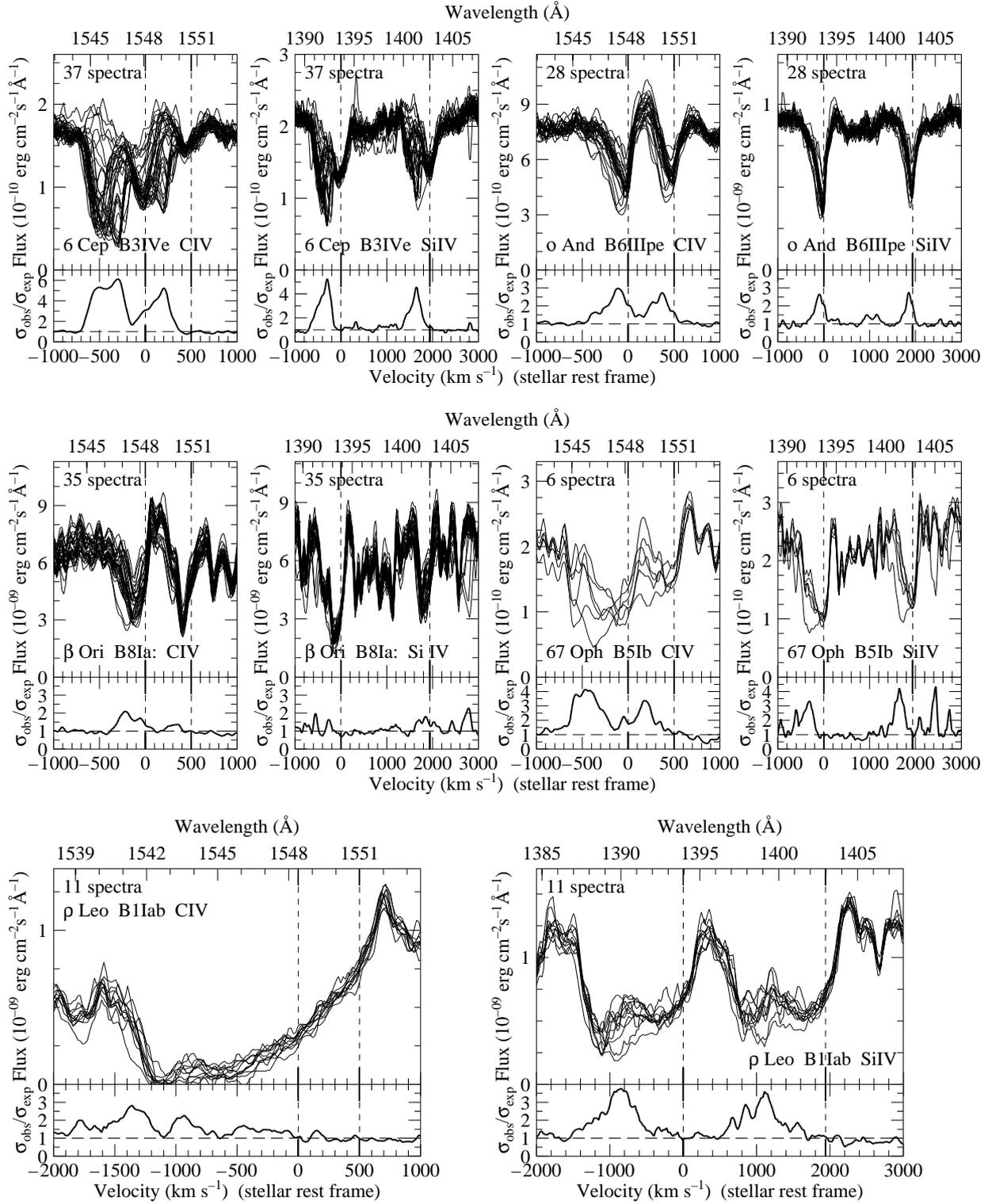}
\caption[]{Same as Fig.~\ref{IUEb1}, but for the second part of the B stars, for which the \ion{Si}{iv} doublet is also shown.}
\label{IUEb2}
\end{center}
\end{figure*}

\begin{figure*}[thbp]
\begin{center}
\includegraphics[width=0.9\linewidth,angle=0,trim=0 0 0 0, clip]{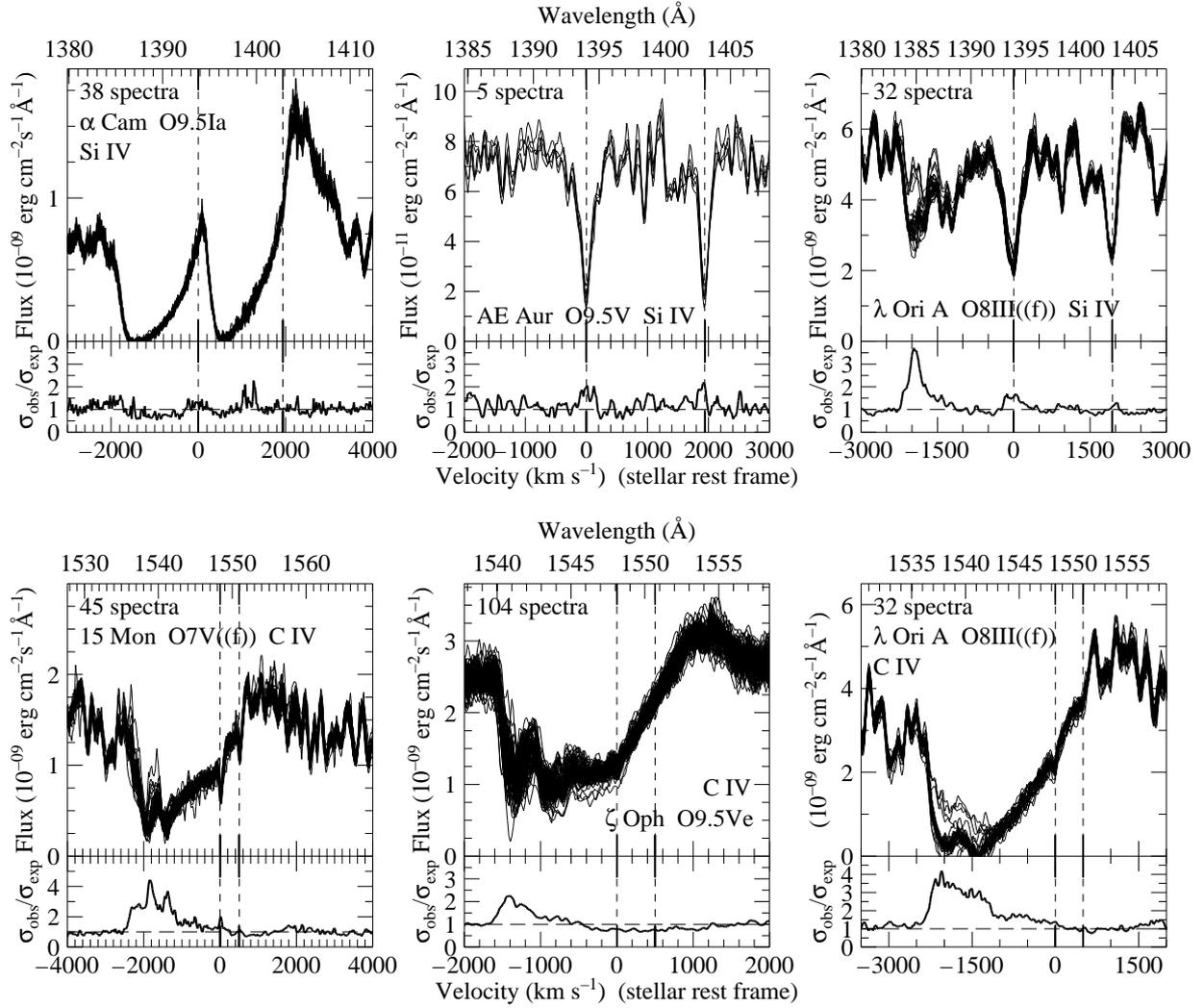}
\caption[]{Same as Figs.~\ref{IUEb1} and \ref{IUEb2}, but for the O stars in our sample. }
\label{IUEo}
\end{center}
\end{figure*}

{\sl HD 203467 (\object{6 Cep})} B3IVe. This is one of the few Be stars in our sample. The emission in  H$\alpha$  and in other lines of this star was studied by \citet{saad:2006}. The UV wind lines of \ion{C}{iv}, \ion{Si}{iv} (see Fig. \ref{IUEb2}), and the \ion{Al}{iii} $\lambda$1855 doublet (not shown) exhibit strong variability of the intermediate type as described in Sect.~\ref{windvariability}, which makes this target a strong candidate.

{\sl HD 207330 (\object{$\pi^2$ Cyg})} B3III. \citet{gies:1992} reported N enhancement for this star. The two IUE spectra were taken 4 hours apart, and are not significantly different. We note that the \ion{C}{iv} profile shows only absorption, and no additional emission, as observed for $\xi^1$~CMa (Fig.~\ref{bcepxi1cma}).

{\sl HD 217675 (\object{o And})} B6IIIpe+A2p. This well-known Be-shell star was reported to be part of a quadruple system \citep{olevic:2006}, with its closest companion, a $\sim$3 M$_{\odot}$ star, in an moderately eccentric 33-day orbit. 
The vicinity to the star of its closest companion could affect its stellar wind. The displayed \ion{C}{iv} and \ion{Si}{iv} (and also the \ion{Al}{iii} $\lambda$1855 doublet, not shown) line-profile variations, are similar to those of magnetic rotators.

{\sl HD 218376 (\object{1 Cas})} B0.5IV. N-enhancement was reported by \citet{gies:1992}. The only two reliable IUE spectra were taken within 2 hours, and no variability is apparent.

\subsubsection{B supergiants}
{\sl HD 34085 (\object{$\beta$ Ori})} B8Ia. DACs were reported by \citet{halliwell:1988} and \citet{bates:1990}, in particular within the UV \ion{Mg}{ii} $\lambda$2800 resonance doublet, which is not shown. The \ion{C}{iv} profiles show similar behaviour.

{\sl HD 91316 (\object{$\rho$ Leo})} B1Iab. \citet{morel:2004} searched for rotationally-modulated H$\alpha$ profiles in this star but did not detect any periodicity between 4.9 and 21.3 days. The snapshot UV wind line profiles clearly show the presence of DACs, but no timeseries data are available for this star.

{\sl HD 164353 (\object{67 Oph})} B5Ib. \citet{koen:2002} reported a 2.3 d period in Hipparcos photometry. The 6 available UV wind spectra contain DACs in all resonance lines.

\subsubsection{O stars}
{\sl HD 30614 (\object{$\alpha$ Cam})}  O9.5Ia. The UV resonance lines are all saturated and show no variability \citep{kaper:1996}, but, in contrast, H$\alpha$ spectra of this runaway star, acquired simultaneously, show rapid variability in the emission \citep{kaper:1997}. \citet{kaper:1997} found that H$\alpha$ emission changes were accompanied by DAC variations for most of the ten O stars included in their study. \citet{crowther:2006} found a systematic N enhancement (in particular N/C) for all studied OB supergiants, including $\alpha$ Cam. \citet{markova:2002} reported on rotationally-modulated wind perturbations. \citet{prinja:2006} investigated wind and atmospheric covariability, and found a possible 0.34d non-radial pulsation in the He $\lambda$5876 line.

{\sl HD 34078 (\object{AE Aur})} O9.5V. This famous runaway star has probably experienced an early dynamical interaction, with both the runaway O9.5V star $\mu$ Col and the O star binary $\iota$ Ori \citep[][ and references therein]{gualandris:2004}; it therefore has a different history from other O stars. CNO abundances were determined by \citet{villamariz:2002}. No obvious UV variability is observed within the 5 available spectra.

{\sl HD 36861 (\object{$\lambda$ Ori A})} O8III((f)). The progression of DACs in the UV resonance lines were studied by \citet{kaper:1996} and \citet{kaper:1999}, but H$\alpha$ profiles, which were measured using data acquired simultaneously, do not show evidence of variations. The \ion{C}{iv} profiles almost reach saturation.

{\sl HD 47839 (\object{15 Mon})} O7V((f)).
Two sets of migrating DACs in the \ion{N}{v} doublet were reported by
\citet{kaper:1996}, who measured their properties and was able to set a lower
limit of 4.5 days to the recurrence timescale. \citet{walborn:2006} noted
that the UV spectra of this star have unexplained peculiarities, a
property shared with the magnetic stars HD191612, $\tau$ Sco and $\xi^1$ CMa.

{\sl HD 149757 (\object{$\zeta$ Oph})} O9.5Vnn. \citet{villamariz:2005} found N enrichment in this very rapidly-rotating runaway star. The UV resonance lines show multiple DACs, which were thoroughly investigated by \citet{howarth:1993}; optical spectroscopic pulsation studies, using simultaneously obtained data, were completed by \citet{reid:1993}.

{\sl HD 214680 (\object{10 Lac})} O9V.  Quantitative measurements of DACs completed for timeseries data acquired in November 1992, are reported in \citet{kaper:1996} and \citet{kaper:1999}. The progression of the DACs in \ion{N}{v} in August
1995, is illustrated in Fig.~\ref{IUE10lac}. A Fourier analysis of this
dataset yielded a period of $6.8\pm1.0$ d. The CNO abundances were
determined by \citet{villamariz:2002}.

\subsubsection{Other targets}
{\sl HD 182989 (\object{RR Lyrae})} F5. This star was added to our target list because this well-known pulsator has an unexplained modulation of the pulsation amplitude \citep[the Blazhko effect, see][]{blazhko:1907}, which could be due to magnetic fields.

\section{Observations \& data reduction}
\label{obs&data}

We observed our target stars and two magnetic calibrators (see Table \ref{targetlist}) using the Musicos spectropolarimeter mounted on the 2m Telescope Bernard Lyot (TBL) at the Pic du Midi, France. The method used to carry out high-precision magnetic field measurements with this instrument is extensively described by \citet{donati:1997} and \citet{wade:2000}. A total of 460 spectra were obtained between December 1998 and June 2003, with a spectral resolution of $R\simeq$35000 within the wavelength range between 449 nm and 662 nm. For each measurement of the effective magnetic-field strength, a set of four subsequent exposures was used. These are acquired in the usual $\lambda/4$-plate position sequence $-$45$^\circ$, +45$^\circ$, +45$^\circ$, $-$45$^\circ$.
We used the dedicated ESpRIT data-reduction package \citep{donati:1997} to complete the optimal extraction of \'{e}chelle orders.
The package includes a Least-Squares Deconvolution (LSD) routine to calculate a high S/N, average Stokes $I$ line profile, and corresponding Stokes $V$ line profile, using all available, magnetically-sensitive, spectral lines.

\subsection{Determining the spectral properties}
To properly combine all available lines from a spectrum using the LSD method, accurate measurements of line depths are required. We determined the depths of lines by fitting the following function to the highest S/N spectrum for each star:

\begin{equation} \label{eq:fitfunc}
F(d_{1..N},\lambda_{0,1..N},\Delta,v_{rad})=1-\sum_{i=1}^{N} d_{i} \mathrm{exp}(-\left(\frac{(1-\frac{\lambda}{\lambda_{0,i}})c-v_{rad}}{\Delta}\right)^2),
\end{equation}
where $N$ is the number of lines in the spectrum, each line of which has a line number $i$, a rest wavelength of $\lambda_{0,i}$, and a central line-depth of $d_{i}$; $\Delta$ is the velocity step, within which the line depth decreases by a factor of $e$; $v_{rad}$ is the radial velocity of the star; and $c$ is the speed of light. The fits were performed using a Levenberg-Marquardt $\chi$-squared minimalisation scheme. The hydrogen lines were excluded from this analysis because they have a different shape (due mostly to Stark broadening) and can therefore not be used in the LSD method.

For stars that had strongly asymmetric lines detected in their spectra, we used $\Delta'$ instead of $\Delta$, where $\Delta'=\Delta+a$ for $\lambda>(1+v_{rad}/c)\lambda_0$; $\Delta'=\Delta-a$, for $\lambda<(1+v_{rad}/c)\lambda_0$; and $a$ is a new parameter that describes the asymmetry of the lines. All spectra with a measurement of $a/\Delta \gtrsim 10\%$ were considered to have strongly asymmetric lines.

For the magnetic calibration stars 53 Cam and $\alpha^2$ CVn, and the F5 star RR Lyrae, we used theoretical line lists and line depths.

\subsection{Measuring the effective magnetic fields}
In combining lines using the LSD method, we assigned each line a weight of $\lambda_{\mathrm{i}} \cdot d_{\mathrm{i}} \cdot g_{\mathrm{eff, i}}$, i.e.\ the product of the wavelength, depth, and effective Land\'{e} factor of the line. Using the average line profiles, we calculated the effective longitudinal field strength $B_l$, as
\begin{equation} \label{eq:Bintegral}
B_l=2.14 \times 10^{11} \frac{\int vV(v)\mathrm{d}v}{\lambda g_\mathrm{av} c \int [1-I(v)]\mathrm{d}v},
\end{equation}
\citep[see ][]{mathys:1989}, where $B_l$ is in gauss, $v$ is the velocity relative to the line centre, $V(v)$ and $I(v)$ are the average Stokes $V$ and $I$ profiles, $\lambda$ and $g_{\mathrm{av}}$ are the average wavelength and Land\'{e} factor of all the lines used in the analysis respectively, and $c$ is the velocity of light in cm~s$^{-1}$.

The integration limits for the measurements of $B_l$ are determined by the line width, which is measuredusing a fit to the spectral lines. For the spectra of stars with symmetric lines, we used integration limits of $[-2\Delta,+2\Delta]$, with the minimum of the Stokes $I$ profile shifted to zero velocity. These limits were chosen to ensure that more than $\sim$95\% of the signal was included in Eq.~\ref{eq:Bintegral}, and to minimise the noise that results from including bins with little or no signal. We verified that using a larger integration range does not significantly alter the results. For the magnetic calibrators 53Cam and $\alpha^2$CVn, the differences were slightly larger. These relatively late type stars, compared to our targets, have many more spectral lines, which implies that the normalisation of the continuum of the average Stoke $I$ profile becomes more difficult. If the integration range is varied, systematic changes will result in the EW, and therefore in the magnetic field measurements. For stars with variable Stokes $I$ profiles and asymmetric lines, we determined the limits by fitting the Stokes $I$ profile resulting from the LSD. The range that we used is $[-2\Delta',+2\Delta']$, where $\Delta'$ is defined as before, and the limits are set by measurements for the most extreme spectra.

\section{Results}
\label{sect:results}

\subsection{The magnetic calibrators}
\begin{figure}[bthp]
\centering
\includegraphics[height=0.95\linewidth,angle=-90,trim=60 10 10 80, clip]{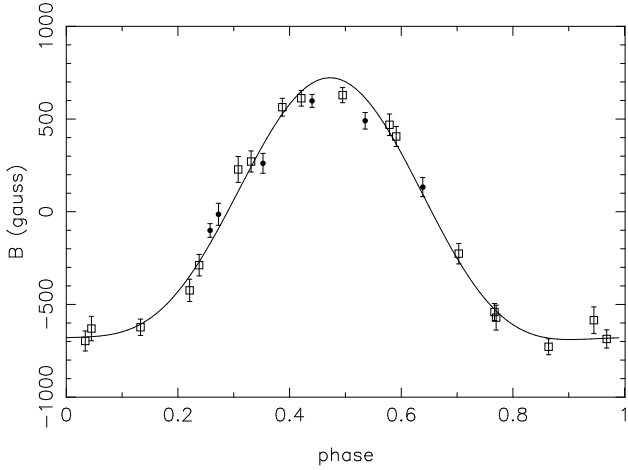}
\caption[]{Longitudinal magnetic-field measurements as a function of phase for $\alpha^2$ CVn (full circles), one of our magnetic calibrator stars, compared with measurements of \citet[][ open squares]{wade:2000}. The line is a second-order Fourier fit to the points of \citet{wade:2000}.}
\label{alfa2cvn}
\end{figure}

\begin{figure}[b!tph]
\centering
\includegraphics[height=0.95\linewidth,angle=-90,trim=60 10 10 80, clip]{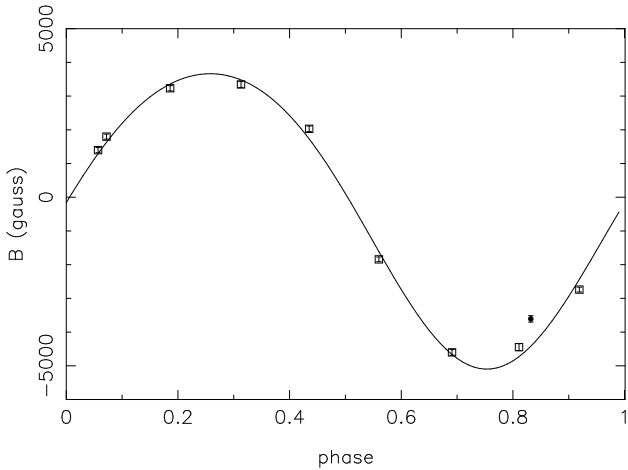}
\caption[]{Same as Fig.\,\ref{alfa2cvn}, but for 53 Cam.}
\label{53cam}
\end{figure}

To test the instrument and establish the orientation of the optics, which determines the sign of the polarisation measured, we included observations of well-known magnetic stars. In Fig.~\ref{alfa2cvn} and \ref{53cam}, we show the longitudinal magnetic-field strength of our magnetic calibrators  $\alpha^2$ CVn and  53 Cam against magnetic (rotational) phase, and compare these data with earlier measurements by \citet{wade:2000}. The ephemeris and period are from \citet[][ $\alpha^2$ CVn]{farnsworth:1932} and \citet[][ 53\,Cam]{hill:1998}. Although the results agree, in general, with the earlier measurements, which confirms the correct operation of the instrument, there are statistically significant deviations in the field strength. These can be due to real changes on the stellar surface, e.g.\ changing abundance patterns \citep[as suggested by][]{wade:2000}, but are most likely related to small differences in the line lists used.

\subsection{Magnetic field measurements}
The derived magnetic-field strengths of the observed targets are listed in Table~\ref{Bfieldslist} (available in the online edition). For a number of targets, a fringe correction was applied (see below).
The targets that were observed most extensively or deserve further comment are discussed below in the same order as they appear in the table.
For the remaining targets, no evidence for the presence of magnetic fields was found, i.e.\ no magnetic-field strengths were measured at a detection level of more than 3$\sigma$.  We do not find any significant circular polarisation signatures in the Stokes $V$ profiles of these targets, which would indicate the presence of a magnetic field. Unfortunately the rotation periods of the observed stars are not known to sufficient precision and the observations are too sparsely-sampled to enable analysis of the magnetic field as a function of rotational phase.

Another quantity that can help to determine the presence of a magnetic field is the weighted-averaged field, $\langle B_{\rm av}\rangle$, and its corresponding error $\langle \sigma_{\rm av}\rangle $ for a set of measurements:

\begin{equation}\label{Bav}
\langle B_{\rm av} \rangle \equiv
\frac{\sum_{i=1}^{n}B_i / \sigma_i^2}{\sum_{i=1}^{n}1/ \sigma_i^2}
\end{equation}
and 

\begin{equation}\label{sigav}
\langle \sigma_{\rm av}\rangle \equiv \sqrt{\sum_{i=1}^{n}1/\sigma_i^2}
\end{equation}
where $B_i$ is the measured value with error $\sigma_i$ of measurement $i$, and $n$ is the total number of observations.
If a series of measurements yields $\langle B_{\rm av}\rangle \gg \langle \sigma_{\rm av}\rangle $, a field is likely to be present. The opposite is, of course, not true: if the average value equals zero, it does not imply the absence of a field, because the configuration can be symmetric.
As an example, for our 6 values for $\alpha^2$ CVn, we obtain $\langle B_{\rm av}\rangle = $260 G and $\langle\sigma_{\rm av}\rangle =$18 G, which confirms that a field is detected.

\subsubsection{B stars}
\begin{figure}[t!b]
\centering
\includegraphics[width=0.95\linewidth,angle=0,clip]{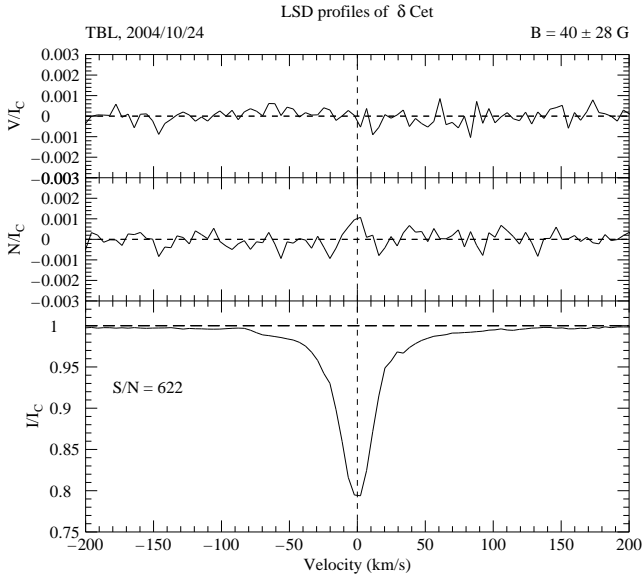}
\caption[]{Results for $\delta$ Cet of 2003/10/24. Average intensity profile ({\bf bottom}) and the Stokes $N$ ({\bf middle}) and $V$ profile ({\bf top}). No magnetic signature is present.}
\label{deltaCet_lsd}
\end{figure}

{\bf $\delta$ Cet}.
As summarised in Sect.\ \ref{targetselection}, this star is one of the strongest magnetic candidates in our sample of B stars. Only one reliable measurement could be obtained of $40\pm28$ G and no magnetic signature in the Stokes $V$ profile was found (see Fig.\ \ref{deltaCet_lsd}). Given the pole-on view of this slow rotator (2 or 4 weeks period), it may be difficult to detect a field, particularly if the angle between the rotation and magnetic axes is close to 90$^{\circ}$, which is the case for a number of other magnetic B stars (see Table \ref{magnetic}). In such a configuration, no rotational modulation can be expected, and the intensity-averaged magnetic field over the entire visible hemisphere, which is the quantity we measure, tends to vanish.\\

\begin{figure}[tb!hp]
\centering
\includegraphics[width=0.95\linewidth,angle=0]{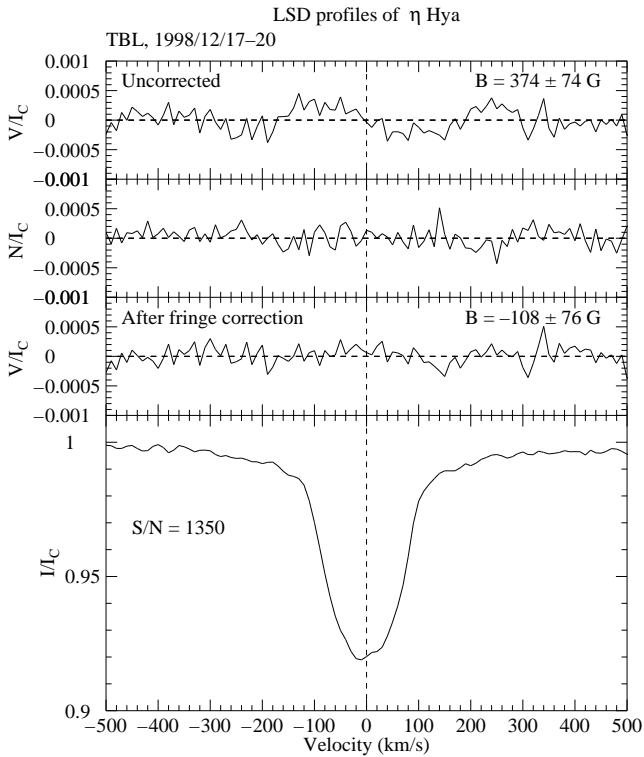}
\caption[]{Weighted average of the three $\eta$ Hya LSD profiles. The average intensity profile ({\bf bottom}) and the Stokes $V$ ({\bf top}) and $N$ profiles before the fringe correction, and the Stokes $V$ profile after the fringe correction ({\bf third panel}) are shown.  A marginal magnetic signature is present without correction for the fringes, which disappears, however, after the correction has been applied.}
\label{etaHya_lsd}
\end{figure}

\noindent{\bf $\eta$ Hya.}
The only B star for which we have a significant magnetic-field detection is $\eta$ Hya. The three observations that we have acquired over a period of three days give a weighted average of $374\pm74$ G (using Eqs.~\ref{Bav}, \ref{sigav}). This is a 5$\sigma$ result, but close inspection of the images shows that fringing on the CCD may have impacted this conclusion. In Fig.~\ref{etaHya_lsd},
we show the average Stokes $V$ profile of these observations, and although a clear signature can be seen in Stokes $V$ at the position of the line, the continuum also appears to be affected by a typical modulation over 250 km s$^{-1}$ intervals. To test whether fringing has indeed affected our measurement, we used observations of the non-magnetic star Vega from the same run to correct for the fringing of the spectra. The procedure we used to remove fringes is discussed in more detail by \citet{eva:2003}.

After the correction was applied, no evidence for a magnetic signature in the Stokes $V$ profile remains. We therefore conclude that the Stokes $V$ signature in this star, resembling a magnetic field signature, is a result of fringing.\\

\noindent{\bf $\tau$ Her.}
We measure two detections at a significance level larger than $3\sigma$: 506$\pm$161 G on 2002/06/12 and 390$\pm$124 G on 2003/06/16. These are, however, not confirmed by measurements from the same or adjacent nights. Applying Eqs.\ \ref{Bav} and \ref{sigav} for the whole dataset of 35 values, we find 23$\pm$ 26 G, which is entirely consistent with a null result.\\

\noindent{\bf 6 Cep.}
This Be star is a strong magnetic candidate because of its wind behaviour (see Sect.~\ref{targetselection}), but only one measurement with a large error bar could be obtained: $1518\pm766$ G, due to the large value of $v \sin i$, which prohibits a more accurate measurement.\\

\noindent{\bf $\pi^2$ Cyg.}
For the set of 15 measurements, we obtain a weighted average of $\langle B_{\rm av}\rangle = -$33 G and a corresponding error $\langle \sigma_{\rm av}\rangle =$30 G (Eqs.\ \ref{Bav}, \ref{sigav}), which is consistent with zero. None of the individual values are significant with typical errors of $\sim 120$ G, and hence no field has been detected.\\

\noindent{\bf $o$ And.}
Our one measurement of this star of $B_\mathrm{eff}=331\pm 988$ does not place strong constraints on its magnetic field. Being a strong magnetic candidate because of its wind emission and modulation in spite of its late spectral type (B6), this star will likely remain a difficult target because of its high value of $v \sin i$ and possible contamination by its companion.\\

\noindent{\bf 1 Cas.}
A weighted average of $\langle B_{\rm av}\rangle = -$1 G with a corresponding error $\langle \sigma_{\rm av}\rangle =$18 G (Eqs.\ \ref{Bav}, \ref{sigav}) is consistent with zero. Among the 18 measurements none of the individual values are significant, and hence no field has been detected.

\subsubsection{O stars}

{\bf 15 Mon.}
No significant detections were found for 5 measurements. This star,however, remains a strong magnetic candidate in view of
its similarities with other magnetic stars as discussed in Sect.~\ref{targetselection}.\\

\begin{figure}[t!hbp]
\centering
\includegraphics[width=0.95\linewidth,clip]{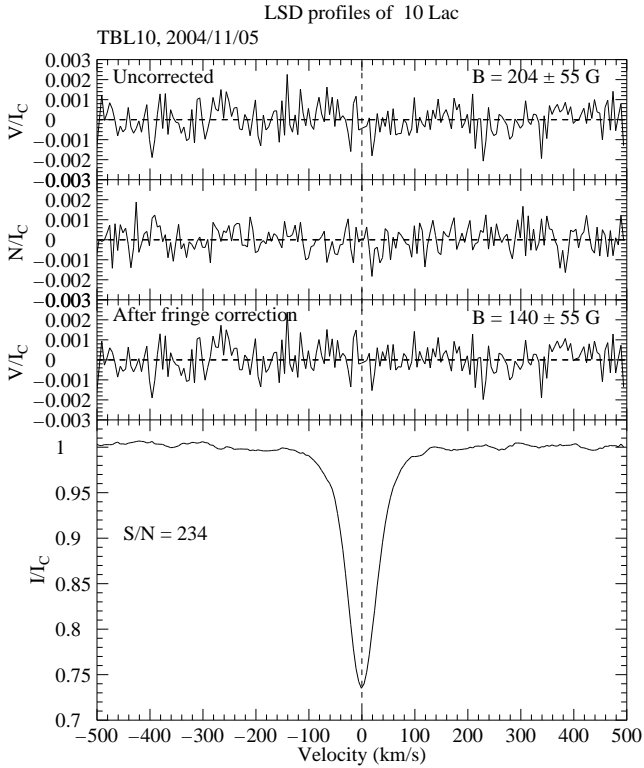}
\caption[]{The LSD intensity profile ({\bf bottom}) and the Stokes $V$ ({\bf top}) and $N$ ({\bf second panel}) profiles of 10 Lac on 2004/11/05 (number 10 in Table \ref{Bfieldslist}). The Stokes $V$ profile after fringe correction is shown in the third panel. The integration limits for the calculation of the magnetic field were [-83, 83] km s$^{-1}$.}
\label{10lac10_lsd}
\end{figure}

\noindent{\bf 10 Lac.}
Although no clear Stokes $V$ signature is found among the 15 individual magnetic-field determinations of this O star, there is one possible significant detection at the 3.7$\sigma$ level of 204 $\pm$ 55 G (see Fig.~\ref{10lac10_lsd}). In addition, we find $\langle B_{\rm av}\rangle = 44\pm14$ G.

We investigated the possible effects of fringes on the magnetic-field measurements, even though no clear modulation, as observed for $\eta$ Hya is evident in the Stokes $V$ spectra of 10 Lac. Assuming that the fringe patterns are similar to those in $\eta$ Hya, Vega and $\beta$ Cep \citep[see][]{henrichs:2006} we find that after applying a fringe correction, the magnetic field values on average decrease by 59 G, which implies that $\langle B_{\rm av}\rangle = -18\pm14$ G. In Fig.~\ref{10lac10_lsd}, we show the Stokes $V$ and $I$ LSD profiles of measurement number 10 in Table \ref{Bfieldslist}, both before and after application of the fringe correction. The 
3.7$\sigma$ detection of 204 $\pm$ 55 G decreases to a value of 140 $\pm$ 55 G, or 2.5$\sigma$.

Whether this result is more reliable than without a correction for fringing is unclear because we are uncertain whether the fringe pattern in the spectra of 10 Lac is similar to that of the fringe template used for the correction, in particular because the noise is relatively high. The cause of the fringes is not completely understood, so the fringes in 10 Lac could well be much weaker than in the spectra of much brighter targets where fringes are usually seen.

It is clear that the evidence for the presence of a magnetic field in 10 Lac depends strongly on our assumptions concerning the fringes and it is premature to claim a detection. Only a careful study of the instrumental effects responsible for the fringes, or new, higher signal-to-noise, observations with new instruments such as Narval at the TBL could provide a more definite answer.

In view of the long recurrence timescale of the DACs (about 1 week or a multiple thereof) and the low value of $v \sin i$ of 31 km s$^{-1}$, the star is most likely a slow rotator that is being viewed almost equator-on.

\subsubsection{Other targets}
{\bf RR Lyrae.}
Our upper limits for the magnetic-field strength of RR Lyrae confirm the results of \citet{chadid:2004}, who did not detect a magnetic field using 27 measurements with typical errors of the order of 70 gauss.

\subsection{Radial velocities and pulsations}
For all of our observations, we measured the heliocentric radial velocity of the minimum in the Stokes $I$ line profile, $v_{min}$, and the first moment of the Stokes $I$ line profile $v_{m1}$; see Table~\ref{Bfieldslist}.

Of all of our targets, four are listed in the 9$^{th}$ Catalogue of spectroscopic binary orbits \citep{pourbaix:2004}. The single-lined binary  $\iota$ Her shows radial-velocity changes in our spectra. $\alpha$ Vir shows both a large radial velocity and an asymmetric line profile that are most likely due to its close double-lined binary nature. We are unable to recognise as binaries using our observations, the single-lined $o$ And, for which we only have one spectrum, and  $\pi^2$ Cyg, which has a small radial-velocity amplitude of 7.8 km s$^{-1}$. RS Sex, BW Vul, $\tau$ Her and RR Lyrae show radial-velocity changes that are most likely due to pulsations. For 67 Oph and 1 Cas, small changes in the radial velocity are observed, which could be an indication of binarity.

\section{Conclusions}
\label{conclusions}
In our survey of 25 OB type stars, we find no conclusive evidence for magnetic fields in B type stars. A possible magnetic-field detection in the O9V star 10 Lac remains uncertain, because the magnetic field values depend critically on the applied correction for fringe effects in the Stokes $V$ spectra. Only with detailed knowledge of the instrumental origin of the fringes improvement can be achieved.
Although for some rapid rotators the error bars are too large to constrain realistic fields, for the majority of the targets, the error bars are of the order of 100 G or smaller. Similar results were obtained in a large survey of B-type stars in open clusters and associations by \citet{bagnulo:2006}. Although the effective field strength of course depends on the orientation of the rotation and magnetic axes, we conclude from these results that strong ($\gtrsim$500 G) fields are certainly not widespread among normal (non chemically peculiar) B-type stars.

It is still possible that the UV wind-line variability, which is observed in a significant fraction of the OB stars \citep[see][]{henrichs:2005}, is due to large-scale magnetic fields. However, such fields would have to be of the order of fifty to a few hundred gauss to remain undetected in these surveys, but still have sufficient impact on the stellar winds. Another possibility is that the perturbation of the wind at the stellar surface \citep[as modelled by ][]{cranmer:1996} is due to strongly magnetic spots. Since these spots cover only a small part of the stellar disk, the local magnetic fields can be quite strong, but still remain undetected. The finite lifetime of such spots would also explain why the UV line-variability has a timescale similar to the rotation period, but is not strictly periodic over longer timescales. If the O star 10 Lac was found to be magnetic, this would provide a strong argument in favour of one of these hypotheses.

We identify a number of stars suitable for follow-up studies: the B stars $\delta$ Cet and 6 Cep, and a number of O stars. In addition, we find excess emission in UV-wind lines centred about the rest wavelength, as observed in $\beta$ Cep and $\xi^1$ CMa, to be a new indirect indicator for the presence of a magnetic field in early B-type stars.

{\acknowledgements  This research is partly based on INES data from the IUE satellite. We would like to thank the helpful staff of the Telescope Bernard Lyot (TBL). Based on data obtained with the TBL at the Observatoire du Pic du Midi (France). This research has made use of NASA's Astrophysics Data System, the SIMBAD database operated at CDS, Strasbourg, France, and the Vienna Atomic Line Database, operated at Institut f\"{u}r Astronomie, Vienna, Austria.}

\bibliographystyle{aa}
\bibliography{./references}

\begin{thebibliography}{106}
\expandafter\ifx\csname natexlab\endcsname\relax\def\natexlab#1{#1}\fi

\bibitem[{{Abt} {et~al.}(2002){Abt}, {Levato}, \& {Grosso}}]{abt:2002}
{Abt}, H.~A., {Levato}, H., \& {Grosso}, M. 2002, \apj, 573, 359

\bibitem[{{Abt} \& {Morrell}(1995)}]{abt:1995}
{Abt}, H.~A. \& {Morrell}, N.~I. 1995, \apjs, 99, 135

\bibitem[{{Adelman}(1992)}]{adelman:1992}
{Adelman}, S.~J. 1992, \mnras, 258, 167

\bibitem[{{Adelman} {et~al.}(2006){Adelman}, {Caliskan}, {Gulliver}, \&
  {Teker}}]{adelman:2006}
{Adelman}, S.~J., {Caliskan}, H., {Gulliver}, A.~F., \& {Teker}, A. 2006, \aap,
  447, 685

\bibitem[{{Aerts} {et~al.}(2006){Aerts}, {Marchenko}, {Matthews}, {Kuschnig},
  {Guenther}, {Moffat}, {Rucinski}, {Sasselov}, {Walker}, \&
  {Weiss}}]{aerts:2006}
{Aerts}, C., {Marchenko}, S.~V., {Matthews}, J.~M., {et~al.} 2006, \apj, 642,
  470

\bibitem[{{Bagnulo} {et~al.}(2006){Bagnulo}, {Landstreet}, {Mason}, {Andretta},
  {Silaj}, \& {Wade}}]{bagnulo:2006}
{Bagnulo}, S., {Landstreet}, J.~D., {Mason}, E., {et~al.} 2006, \aap, 450, 777

\bibitem[{{Bates} \& {Gilheany}(1990)}]{bates:1990}
{Bates}, B. \& {Gilheany}, S. 1990, \mnras, 243, 320

\bibitem[{{Bergh\"{o}fer} {et~al.}(1996){Bergh\"{o}fer}, {Schmitt}, \&
  {Cassinelli}}]{berghofer:1996}
{Bergh\"{o}fer}, T.~W., {Schmitt}, J.~H.~M.~M., \& {Cassinelli}, J.~P. 1996,
  \aaps, 118, 481

\bibitem[{{Bla{\v z}ko}(1907)}]{blazhko:1907}
{Bla{\v z}ko}, S. 1907, Astronomische Nachrichten, 175, 325

\bibitem[{{Braithwaite} \& {Spruit}(2004)}]{braithwaite:2004b}
{Braithwaite}, J. \& {Spruit}, H.~C. 2004, \nat, 431, 819

\bibitem[{{Briquet} {et~al.}(2003){Briquet}, {Aerts}, {Mathias}, {Scuflaire},
  \& {Noels}}]{briquet:2003}
{Briquet}, M., {Aerts}, C., {Mathias}, P., {Scuflaire}, R., \& {Noels}, A.
  2003, \aap, 401, 281

\bibitem[{{Bychkov} {et~al.}(2003){Bychkov}, {Bychkova}, \&
  {Madej}}]{bychkov:2003}
{Bychkov}, V.~D., {Bychkova}, L.~V., \& {Madej}, J. 2003, \aap, 407, 631

\bibitem[{{Cassinelli}(1985)}]{cassinelli:1985}
{Cassinelli}, J.~P. 1985, {Evidence for non-radiative activity in hot stars},
  Tech. rep., {NASA}

\bibitem[{{Catala} {et~al.}(2007){Catala}, {Alecian}, {Donati}, {Wade},
  {Landstreet}, {B{\"o}hm}, {Bouret}, {Bagnulo}, {Folsom}, \&
  {Silvester}}]{catala:2007}
{Catala}, C., {Alecian}, E., {Donati}, J.-F., {et~al.} 2007, \aap, 462, 293

\bibitem[{{Chadid} {et~al.}(2004){Chadid}, {Wade}, {Shorlin}, \&
  {Landstreet}}]{chadid:2004}
{Chadid}, M., {Wade}, G.~A., {Shorlin}, S.~L.~S., \& {Landstreet}, J.~D. 2004,
  \aap, 413, 1087

\bibitem[{{Charbonneau} \& {MacGregor}(2001)}]{charbonneau:2001}
{Charbonneau}, P. \& {MacGregor}, K.~B. 2001, \apj, 559, 1094

\bibitem[{{Cohen} {et~al.}(2003){Cohen}, {de Messi{\` e}res}, {MacFarlane},
  {Miller}, {Cassinelli}, {Owocki}, \& {Liedahl}}]{cohen:2003}
{Cohen}, D.~H., {de Messi{\` e}res}, G.~E., {MacFarlane}, J.~J., {et~al.} 2003,
  \apj, 586, 495

\bibitem[{{Cranmer} \& {Owocki}(1996)}]{cranmer:1996}
{Cranmer}, S.~R. \& {Owocki}, S.~P. 1996, \apj, 462, 469

\bibitem[{{Crowther} {et~al.}(2006){Crowther}, {Lennon}, \&
  {Walborn}}]{crowther:2006}
{Crowther}, P.~A., {Lennon}, D.~J., \& {Walborn}, N.~R. 2006, \aap, 446, 279

\bibitem[{{Crutcher}(1999)}]{crutcher:1999}
{Crutcher}, R.~M. 1999, \apj, 520, 706

\bibitem[{{de Jong} {et~al.}(2001){de Jong}, {Henrichs}, {Kaper}, {Nichols},
  {Bjorkman}, {Bohlender}, {Cao}, {Gordon}, {Hill}, {Jiang}, {Kolka},
  {Morrison}, {Neff}, {O'Neal}, {Scheers}, \& {Telting}}]{dejong:2001}
{de Jong}, J.~A., {Henrichs}, H.~F., {Kaper}, L., {et~al.} 2001, \aap, 368, 601

\bibitem[{{de Jong} {et~al.}(1999){de Jong}, {Henrichs}, {Schrijvers}, {Gies},
  {Telting}, {Kaper}, \& {Zwarthoed}}]{dejong:1999}
{de Jong}, J.~A., {Henrichs}, H.~F., {Schrijvers}, C., {et~al.} 1999, \aap,
  345, 172

\bibitem[{{Donati} {et~al.}(2002){Donati}, {Babel}, {Harries}, {Howarth},
  {Petit}, \& {Semel}}]{donati:2002}
{Donati}, J.-F., {Babel}, J., {Harries}, T.~J., {et~al.} 2002, \mnras, 333, 55

\bibitem[{{Donati} {et~al.}(2006{\natexlab{a}}){Donati}, {Howarth}, {Bouret},
  {Petit}, {Catala}, \& {Landstreet}}]{donati:2006}
{Donati}, J.-F., {Howarth}, I.~D., {Bouret}, J.-C., {et~al.}
  2006{\natexlab{a}}, \mnras, 365, L6

\bibitem[{{Donati} {et~al.}(2006{\natexlab{b}}){Donati}, {Howarth}, {Jardine},
  {Petit}, {Catala}, {Landstreet}, {Bouret}, {Alecian}, {Barnes}, {Forveille},
  {Paletou}, \& {Manset}}]{donati:2006b}
{Donati}, J.-F., {Howarth}, I.~D., {Jardine}, M.~M., {et~al.}
  2006{\natexlab{b}}, \mnras, 370, 629

\bibitem[{{Donati} {et~al.}(1997){Donati}, {Semel}, {Carter}, {Rees}, \&
  {Collier Cameron}}]{donati:1997}
{Donati}, J.-F., {Semel}, M., {Carter}, B.~D., {Rees}, D.~E., \& {Collier
  Cameron}, A. 1997, \mnras, 291, 658

\bibitem[{{Donati} {et~al.}(2001){Donati}, {Wade}, {Babel}, {Henrichs}, {de
  Jong}, \& {Harries}}]{donati:2001}
{Donati}, J.-F., {Wade}, G.~A., {Babel}, J., {et~al.} 2001, \mnras, 326, 1265

\bibitem[{{Farnsworth}(1932)}]{farnsworth:1932}
{Farnsworth}, G. 1932, \apj, 76, 313

\bibitem[{{Ferrario} \& {Wickramasinghe}(2005)}]{ferrario:2005}
{Ferrario}, L. \& {Wickramasinghe}, D.~T. 2005, \mnras, 356, 615

\bibitem[{{Fullerton}(2003)}]{fullerton:2003}
{Fullerton}, A.~W. 2003, in ASP Conf. Ser., Vol. 305, ``Magnetic Fields in O, B
  and A Stars: Origin and Connection to Pulsation, Rotation and Mass Loss'',
  333

\bibitem[{{Fullerton} {et~al.}(1996){Fullerton}, {Gies}, \&
  {Bolton}}]{fullerton:1996}
{Fullerton}, A.~W., {Gies}, D.~R., \& {Bolton}, C.~T. 1996, \apjs, 103, 475

\bibitem[{{Gies} \& {Lambert}(1992)}]{gies:1992}
{Gies}, D.~R. \& {Lambert}, D.~L. 1992, \apj, 387, 673

\bibitem[{{Grigsby} {et~al.}(1996){Grigsby}, {Mulliss}, \&
  {Baer}}]{grigsby:1996}
{Grigsby}, J.~A., {Mulliss}, C.~L., \& {Baer}, G.~M. 1996, \pasp, 108, 953

\bibitem[{{Gualandris} {et~al.}(2004){Gualandris}, {Portegies Zwart}, \&
  {Eggleton}}]{gualandris:2004}
{Gualandris}, A., {Portegies Zwart}, S., \& {Eggleton}, P.~P. 2004, \mnras,
  350, 615

\bibitem[{{Halliwell} {et~al.}(1988){Halliwell}, {Bates}, \&
  {Catney}}]{halliwell:1988}
{Halliwell}, D.~R., {Bates}, B., \& {Catney}, M.~G. 1988, \aap, 189, 204

\bibitem[{{Harnden} {et~al.}(1979){Harnden}, {Branduardi}, {Gorenstein},
  {Grindlay}, {Rosner}, {Topka}, {Elvis}, {Pye}, \& {Vaiana}}]{harnden:1979}
{Harnden}, Jr., F.~R., {Branduardi}, G., {Gorenstein}, P., {et~al.} 1979,
  \apjl, 234, L51

\bibitem[{{Heger} {et~al.}(2005){Heger}, {Woosley}, \& {Spruit}}]{heger:2005}
{Heger}, A., {Woosley}, S.~E., \& {Spruit}, H.~C. 2005, \apj, 626, 350

\bibitem[{{Henrichs}(1999)}]{henrichs:1999}
{Henrichs}, H.~F. 1999, Lecture Notes in Physics, Berlin Springer Verlag, 523,
  305

\bibitem[{{Henrichs} {et~al.}(2000){Henrichs}, {de Jong}, {Donati}, {Catala},
  {Wade}, {Shorlin}, {Veen}, {Nichols}, \& {Kaper}}]{henrichs:2000a}
{Henrichs}, H.~F., {de Jong}, J.~A., {Donati}, J.-F., {et~al.} 2000, in ASP
  Conf. Ser. 214: IAU Colloq. 175: The Be Phenomenon in Early-Type Stars, ed.
  M.~A. {Smith}, H.~F. {Henrichs}, \& J.~{Fabregat}, 324

\bibitem[{{Henrichs} {et~al.}(2008){Henrichs}, {de Jong}, {Verdugo}, {Schnerr},
  {Neiner}, {Donati}, {Catala}, {Shorlin}, \& {et al.}}]{henrichs:2006}
{Henrichs}, H.~F., {de Jong}, J.~A., {Verdugo}, E., {et~al.} 2008, {submitted}

\bibitem[{{Henrichs} {et~al.}(1994){Henrichs}, {Kaper}, \&
  {Nichols}}]{henrichs:1994}
{Henrichs}, H.~F., {Kaper}, L., \& {Nichols}, J.~S. 1994, \aap, 285, 565

\bibitem[{{Henrichs} {et~al.}(2005){Henrichs}, {Schnerr}, \& {ten
  Kulve}}]{henrichs:2005}
{Henrichs}, H.~F., {Schnerr}, R.~S., \& {ten Kulve}, E. 2005, in ASP Conf.
  Ser., Vol. 337: The Nature and Evolution of Disks Around Hot Stars, 114

\bibitem[{{Hill} {et~al.}(1998){Hill}, {Bohlender}, {Landstreet}, {Wade},
  {Manset}, \& {Bastien}}]{hill:1998}
{Hill}, G.~M., {Bohlender}, D.~A., {Landstreet}, J.~D., {et~al.} 1998, \mnras,
  297, 236

\bibitem[{{Hoffleit} \& {Jaschek}(1991)}]{bsc5:1991}
{Hoffleit}, D. \& {Jaschek}, C. 1991, {The Bright Star Catalogue} (New Haven,
  Conn.: Yale University Observatory, 5th rev.ed., edited by D.\ Hoffleit and
  C.\ Jaschek)

\bibitem[{{Houk} \& {Swift}(1999)}]{houk:1999}
{Houk}, N. \& {Swift}, C. 1999, {Michigan catalogue of two-dimensional spectral
  types for the HD Stars, Vol.\ 5} (Michigan: Department of Astronomy,
  University of Michigan)

\bibitem[{{Howarth} {et~al.}(1993){Howarth}, {Bolton}, {Crowe}, {Ebbets},
  {Fieldus}, {Fullerton}, {Gies}, {McDavid}, {Prinja}, {Reid}, {Shore}, \&
  {Smith}}]{howarth:1993}
{Howarth}, I.~D., {Bolton}, C.~T., {Crowe}, R.~A., {et~al.} 1993, \apj, 417,
  338

\bibitem[{{Hubrig} {et~al.}(2006{\natexlab{a}}){Hubrig}, {Briquet},
  {Sch{\"o}ller}, {De Cat}, {Mathys}, \& {Aerts}}]{hubrig:2006}
{Hubrig}, S., {Briquet}, M., {Sch{\"o}ller}, M., {et~al.} 2006{\natexlab{a}},
  \mnras, 369, L61

\bibitem[{{Hubrig} {et~al.}(2004){Hubrig}, {Sch{\"o}ller}, \&
  {Yudin}}]{hubrig:2004}
{Hubrig}, S., {Sch{\"o}ller}, M., \& {Yudin}, R.~V. 2004, \aap, 428, L1

\bibitem[{{Hubrig} {et~al.}(2006{\natexlab{b}}){Hubrig}, {Yudin},
  {Sch{\"o}ller}, \& {Pogodin}}]{hubrig:2006b}
{Hubrig}, S., {Yudin}, R.~V., {Sch{\"o}ller}, M., \& {Pogodin}, M.~A.
  2006{\natexlab{b}}, \aap, 446, 1089

\bibitem[{{Kaper} {et~al.}(1997){Kaper}, {Henrichs}, {Fullerton}, {Ando},
  {Bjorkman}, {Gies}, {Hirata}, {Kambe}, {McDavid}, \& {Nichols}}]{kaper:1997}
{Kaper}, L., {Henrichs}, H.~F., {Fullerton}, A.~W., {et~al.} 1997, \aap, 327,
  281

\bibitem[{{Kaper} {et~al.}(1996){Kaper}, {Henrichs}, {Nichols}, {Snoek},
  {Volten}, \& {Zwarthoed}}]{kaper:1996}
{Kaper}, L., {Henrichs}, H.~F., {Nichols}, J.~S., {et~al.} 1996, \aaps, 116,
  257

\bibitem[{{Kaper} {et~al.}(1999){Kaper}, {Henrichs}, {Nichols}, \&
  {Telting}}]{kaper:1999}
{Kaper}, L., {Henrichs}, H.~F., {Nichols}, J.~S., \& {Telting}, J.~H. 1999,
  \aap, 344, 231

\bibitem[{{Kodaira} \& {Scholz}(1970)}]{kodaira:1970}
{Kodaira}, K. \& {Scholz}, M. 1970, \aap, 6, 93

\bibitem[{{Koen} \& {Eyer}(2002)}]{koen:2002}
{Koen}, C. \& {Eyer}, L. 2002, \mnras, 331, 45

\bibitem[{{Lyubimkov} {et~al.}(2004){Lyubimkov}, {Rostopchin}, \&
  {Lambert}}]{lyubimkov:2004}
{Lyubimkov}, L.~S., {Rostopchin}, S.~I., \& {Lambert}, D.~L. 2004, \mnras, 351,
  745

\bibitem[{{MacDonald} \& {Mullan}(2004)}]{macdonald:2004}
{MacDonald}, J. \& {Mullan}, D.~J. 2004, \mnras, 348, 702

\bibitem[{{Maeder} \& {Meynet}(2003)}]{maeder:2003}
{Maeder}, A. \& {Meynet}, G. 2003, \aap, 411, 543

\bibitem[{{Maeder} \& {Meynet}(2004)}]{maeder:2004}
{Maeder}, A. \& {Meynet}, G. 2004, \aap, 422, 225

\bibitem[{{Maheswaran} \& {Cassinelli}(1992)}]{maheswaran:1992}
{Maheswaran}, M. \& {Cassinelli}, J.~P. 1992, \apj, 386, 695

\bibitem[{{Ma{\'{\i}}z-Apell{\'a}niz}
  {et~al.}(2004){Ma{\'{\i}}z-Apell{\'a}niz}, {Walborn}, {Galu{\'e}}, \&
  {Wei}}]{maiz:2004}
{Ma{\'{\i}}z-Apell{\'a}niz}, J., {Walborn}, N.~R., {Galu{\'e}}, H.~{\'A}., \&
  {Wei}, L.~H. 2004, \apjs, 151, 103

\bibitem[{{Manchester}(2004)}]{manchester:2004}
{Manchester}, R.~N. 2004, Science, 304, 542

\bibitem[{{Markova}(2002)}]{markova:2002}
{Markova}, N. 2002, \aap, 385, 479

\bibitem[{{Mathys}(1989)}]{mathys:1989}
{Mathys}, G. 1989, Fundamentals of Cosmic Physics, 13, 143

\bibitem[{{Mathys}(2001)}]{mathys:2001}
{Mathys}, G. 2001, in ASP Conf. Ser. 248: Magnetic Fields Across the
  Hertzsprung-Russell Diagram, 267

\bibitem[{{Mokiem} {et~al.}(2005){Mokiem}, {de Koter}, {Puls}, {Herrero},
  {Najarro}, \& {Villamariz}}]{mokiem:2005}
{Mokiem}, M.~R., {de Koter}, A., {Puls}, J., {et~al.} 2005, \aap, 441, 711

\bibitem[{{Morel} {et~al.}(2006){Morel}, {Butler}, {Aerts}, {Neiner}, \&
  {Briquet}}]{morel:2006}
{Morel}, T., {Butler}, K., {Aerts}, C., {Neiner}, C., \& {Briquet}, M. 2006,
  \aap, 457, 651

\bibitem[{{Morel} {et~al.}(2004){Morel}, {Marchenko}, {Pati}, {Kuppuswamy},
  {Carini}, {Wood}, \& {Zimmerman}}]{morel:2004}
{Morel}, T., {Marchenko}, S.~V., {Pati}, A.~K., {et~al.} 2004, \mnras, 351, 552

\bibitem[{{Mullan} \& {MacDonald}(2005)}]{mullan:2005}
{Mullan}, D.~J. \& {MacDonald}, J. 2005, \mnras, 356, 1139

\bibitem[{{Neiner} {et~al.}(2003{\natexlab{a}}){Neiner}, {Geers}, {Henrichs},
  {Floquet}, {Fr{\' e}mat}, {Hubert}, {Preuss}, \& {Wiersema}}]{coralie:2003a}
{Neiner}, C., {Geers}, V.~C., {Henrichs}, H.~F., {et~al.} 2003{\natexlab{a}},
  \aap, 406, 1019

\bibitem[{{Neiner} {et~al.}(2003{\natexlab{b}}){Neiner}, {Henrichs}, {Floquet},
  {Fr{\' e}mat}, {Preuss}, {Hubert}, {Geers}, {Tijani}, {Nichols}, \&
  {Jankov}}]{coralie:2003c}
{Neiner}, C., {Henrichs}, H.~F., {Floquet}, M., {et~al.} 2003{\natexlab{b}},
  \aap, 411, 565

\bibitem[{{Neiner} {et~al.}(2003{\natexlab{c}}){Neiner}, {Hubert}, {Fr{\'
  e}mat}, {Floquet}, {Jankov}, {Preuss}, {Henrichs}, \&
  {Zorec}}]{coralie:2003b}
{Neiner}, C., {Hubert}, A.-M., {Fr{\' e}mat}, Y., {et~al.} 2003{\natexlab{c}},
  \aap, 409, 275

\bibitem[{{Niemczura}(2003)}]{niemczura:2003}
{Niemczura}, E. 2003, \aap, 404, 689

\bibitem[{{Olevi{\'c}} \& {Cvetkovi{\'c}}(2006)}]{olevic:2006}
{Olevi{\'c}}, D. \& {Cvetkovi{\'c}}, Z. 2006, \aj, 131, 1721

\bibitem[{{Oskinova} {et~al.}(2004){Oskinova}, {Feldmeier}, \&
  {Hamann}}]{oskinova:2004}
{Oskinova}, L.~M., {Feldmeier}, A., \& {Hamann}, W.-R. 2004, \aap, 422, 675

\bibitem[{{Oskinova} {et~al.}(2006){Oskinova}, {Feldmeier}, \&
  {Hamann}}]{oskinova:2006}
{Oskinova}, L.~M., {Feldmeier}, A., \& {Hamann}, W.-R. 2006, \mnras, 372, 313

\bibitem[{{Peters}(1976)}]{peters:1976}
{Peters}, G.~J. 1976, \apjs, 30, 551

\bibitem[{{Peterson} {et~al.}(1996){Peterson}, {Carney}, \&
  {Latham}}]{peterson:1996}
{Peterson}, R.~C., {Carney}, B.~W., \& {Latham}, D.~W. 1996, \apjl, 465, L47

\bibitem[{{Pintado} \& {Adelman}(1993)}]{pintado:1993}
{Pintado}, O.~I. \& {Adelman}, S.~J. 1993, \mnras, 264, 63

\bibitem[{{Pourbaix} {et~al.}(2004){Pourbaix}, {Tokovinin}, {Batten}, {Fekel},
  {Hartkopf}, {Levato}, {Morrell}, {Torres}, \& {Udry}}]{pourbaix:2004}
{Pourbaix}, D., {Tokovinin}, A.~A., {Batten}, A.~H., {et~al.} 2004, \aap, 424,
  727

\bibitem[{{Prinja} {et~al.}(2006){Prinja}, {Markova}, {Scuderi}, \&
  {Markov}}]{prinja:2006}
{Prinja}, R.~K., {Markova}, N., {Scuderi}, S., \& {Markov}, H. 2006, \aap, 457,
  987

\bibitem[{{Reid} {et~al.}(1993){Reid}, {Bolton}, {Crowe}, {Fieldus},
  {Fullerton}, {Gies}, {Howarth}, {McDavid}, {Prinja}, \& {Smith}}]{reid:1993}
{Reid}, A.~H.~N., {Bolton}, C.~T., {Crowe}, R.~A., {et~al.} 1993, \apj, 417,
  320

\bibitem[{{Reid} \& {Howarth}(1996)}]{reid:1996}
{Reid}, A.~H.~N. \& {Howarth}, I.~D. 1996, \aap, 311, 616

\bibitem[{{Rodr{\'{\i}}guez-Merino} {et~al.}(2005){Rodr{\'{\i}}guez-Merino},
  {Chavez}, {Bertone}, \& {Buzzoni}}]{rodriguez:2005}
{Rodr{\'{\i}}guez-Merino}, L.~H., {Chavez}, M., {Bertone}, E., \& {Buzzoni}, A.
  2005, \apj, 626, 411

\bibitem[{{Saad} {et~al.}(2006){Saad}, {Kub{\'a}t}, {Kor{\v c}{\'a}kov{\'a}},
  {Koubsk{\'y}}, {{\v S}koda}, {{\v S}lechta}, {Kawka}, {Budovi{\v c}ov{\'a}},
  {Votruba}, {{\v S}arounov{\'a}}, \& {Nouh}}]{saad:2006}
{Saad}, S.~M., {Kub{\'a}t}, J., {Kor{\v c}{\'a}kov{\'a}}, D., {et~al.} 2006,
  \aap, 450, 427

\bibitem[{{Schnerr} {et~al.}(2007){Schnerr}, {Henrichs}, {Owocki}, {Ud-Doula},
  \& {Townsend}}]{schnerr:2007}
{Schnerr}, R.~S., {Henrichs}, H.~F., {Owocki}, S.~P., {Ud-Doula}, A., \&
  {Townsend}, R.~H.~D. 2007, in Astronomical Society of the Pacific Conference
  Series, Vol. 361, Active OB-Stars: Laboratories for Stellare and
  Circumstellar Physics, ed. A.~T. {Okazaki}, S.~P. {Owocki}, \& S.~{Stefl},
  488

\bibitem[{{Schnerr} {et~al.}(2006){Schnerr}, {Verdugo}, {Henrichs}, \&
  {Neiner}}]{schnerr:2006a}
{Schnerr}, R.~S., {Verdugo}, E., {Henrichs}, H.~F., \& {Neiner}, C. 2006, \aap,
  452, 969

\bibitem[{{Schulz} {et~al.}(2000){Schulz}, {Canizares}, {Huenemoerder}, \&
  {Lee}}]{schulz:2000}
{Schulz}, N.~S., {Canizares}, C.~R., {Huenemoerder}, D., \& {Lee}, J.~C. 2000,
  \apjl, 545, L135

\bibitem[{{Seward} {et~al.}(1979){Seward}, {Forman}, {Giacconi}, {Griffiths},
  {Harnden}, {Jones}, \& {Pye}}]{seward:1979}
{Seward}, F.~D., {Forman}, W.~R., {Giacconi}, R., {et~al.} 1979, \apjl, 234,
  L55

\bibitem[{{Shore}(1987)}]{shore:1987}
{Shore}, S.~N. 1987, \aj, 94, 731

\bibitem[{{Slettebak} {et~al.}(1975){Slettebak}, {Collins}, {Parkinson},
  {Boyce}, \& {White}}]{slettebak:1975}
{Slettebak}, A., {Collins}, G.~W., {Parkinson}, T.~D., {Boyce}, P.~B., \&
  {White}, N.~M. 1975, \apjs, 29, 137

\bibitem[{{Spruit}(2002)}]{spruit:2002}
{Spruit}, H.~C. 2002, \aap, 381, 923

\bibitem[{{Stankov} {et~al.}(2003){Stankov}, {Ilyin}, \&
  {Fridlund}}]{stankov:2003}
{Stankov}, A., {Ilyin}, I., \& {Fridlund}, C.~V.~M. 2003, \aap, 408, 1077

\bibitem[{{ten Kulve}(2004)}]{evan:2004}
{ten Kulve}, E. 2004, Master's thesis, {University of Amsterdam}

\bibitem[{{ud-Doula} \& {Owocki}(2002)}]{asif:2002}
{ud-Doula}, A. \& {Owocki}, S.~P. 2002, \apj, 576, 413

\bibitem[{{Valenti} \& {Johns-Krull}(2004)}]{valenti:2004}
{Valenti}, J.~A. \& {Johns-Krull}, C.~M. 2004, \apss, 292, 619

\bibitem[{{Verdugo} {et~al.}(2003){Verdugo}, {Talavera}, {G{\'o}mez de Castro},
  {Henrichs}, {Geers}, \& {Wiersema}}]{eva:2003}
{Verdugo}, E., {Talavera}, A., {G{\'o}mez de Castro}, A.~I., {et~al.} 2003, in
  EAS Publications Series, ed. J.~{Arnaud} \& N.~{Meunier}, 271

\bibitem[{{Villamariz} \& {Herrero}(2005)}]{villamariz:2005}
{Villamariz}, M.~R. \& {Herrero}, A. 2005, \aap, 442, 263

\bibitem[{{Villamariz} {et~al.}(2002){Villamariz}, {Herrero}, {Becker}, \&
  {Butler}}]{villamariz:2002}
{Villamariz}, M.~R., {Herrero}, A., {Becker}, S.~R., \& {Butler}, K. 2002,
  \aap, 388, 940

\bibitem[{{Wade} {et~al.}(2000){Wade}, {Donati}, {Landstreet}, \&
  {Shorlin}}]{wade:2000}
{Wade}, G.~A., {Donati}, J.-F., {Landstreet}, J.~D., \& {Shorlin}, S.~L.~S.
  2000, \mnras, 313, 851

\bibitem[{{Wade} {et~al.}(2005){Wade}, {Drouin}, {Bagnulo}, {Landstreet},
  {Mason}, {Silvester}, {Alecian}, {B{\"o}hm}, {Bouret}, {Catala}, \&
  {Donati}}]{wade:2005}
{Wade}, G.~A., {Drouin}, D., {Bagnulo}, S., {et~al.} 2005, \aap, 442, L31

\bibitem[{{Walborn}(1972)}]{walborn:1972}
{Walborn}, N.~R. 1972, \aj, 77, 312

\bibitem[{{Walborn}(1973)}]{walborn:1973}
{Walborn}, N.~R. 1973, \aj, 78, 1067

\bibitem[{{Walborn}(2006)}]{walborn:2006}
{Walborn}, N.~R. 2006, in Proceedings of the Joint Discussion 4 at the IAU
  General Assembly, Prague, August 16-17, 2006, ed. A.~I. {Gomez de Castro} \&
  M.~{Barstow}, {in press}

\bibitem[{{Walborn} \& {Fitzpatrick}(1990)}]{walborn:1990}
{Walborn}, N.~R. \& {Fitzpatrick}, E.~L. 1990, \pasp, 102, 379

\bibitem[{{Wickramasinghe} \& {Ferrario}(2000)}]{wickramasinghe:2000}
{Wickramasinghe}, D.~T. \& {Ferrario}, L. 2000, \pasp, 112, 873

\bibitem[{{Wolff} {et~al.}(2004){Wolff}, {Strom}, \&
  {Hillenbrand}}]{wolff:2004}
{Wolff}, S.~C., {Strom}, S.~E., \& {Hillenbrand}, L.~A. 2004, \apj, 601, 979

\end{thebibliography}

\Online

{\small
\begin{longtable}{@{ }r@{$\:\:\:$}c@{$\:\:\:$}rr@{$\pm$}l@{$\,$}r@{$\pm$}l@{$\,$}r@{$\pm$}l|r@{$\:\:\:$}c@{$\:\:\:$}rr@{$\pm$}l@{$\:\:$}r@{$\pm$}l@{$\:\:$}r@{$\pm$}l}
\caption{Summary of the results of the data analysis. The columns denote, respectively: sequence number of the observation, date of observation, Heliocentric Julian date, measured longitudinal component of magnetic field strength (integrated over the stellar disk) with 1$\sigma$ errors, velocity of the minimum, and first moment of the Stokes $I$ line profile, both with 1$\sigma$ errors, for all observed targets arranged in the same order as in Table \ref{targetlist}.}
\label{Bfieldslist}
\\
\hline
\hline
Obs & Date  &  HJD        & \multicolumn{2}{c}{$B_\mathrm{eff}$} & \multicolumn{2}{c}{$v_\mathrm{min}$} & \multicolumn{2}{c|}{$v_\mathrm{m1}$} & Obs & Date     &  HJD        & \multicolumn{2}{c}{$B_\mathrm{eff}$} & \multicolumn{2}{c}{$v_\mathrm{min}$} & \multicolumn{2}{c}{$v_\mathrm{m1}$}\\
nr  & year/m/d & $-$2451000  & \multicolumn{2}{c}{gauss}         & \multicolumn{2}{c}{km s$^{-1}$}      & \multicolumn{2}{c|}{km s$^{-1}$}     &nr   & year/m/d & $-$2451000  & \multicolumn{2}{c}{gauss}            & \multicolumn{2}{c}{km s$^{-1}$}      & \multicolumn{2}{c}{km s$^{-1}$}\\
\hline
\endfirsthead
\caption{continued.}\\
\hline\hline
Obs & Date  &  HJD        & \multicolumn{2}{c}{$B_\mathrm{eff}$} & \multicolumn{2}{c}{$v_\mathrm{min}$} & \multicolumn{2}{c|}{$v_\mathrm{m1}$} & Obs & Date  &  HJD        & \multicolumn{2}{c}{$B_\mathrm{eff}$} & \multicolumn{2}{c}{$v_\mathrm{min}$} & \multicolumn{2}{c}{$v_\mathrm{m1}$}\\
nr  & year/m/d & $-$2451000  & \multicolumn{2}{c}{(gauss)}       & \multicolumn{2}{c}{(km s$^{-1}$)}    & \multicolumn{2}{c|}{(km s$^{-1}$)}   &nr  & year/m/d & $-$2451000  & \multicolumn{2}{c}{(gauss)} & \multicolumn{2}{c}{(km s$^{-1}$)}& \multicolumn{2}{c}{(km s$^{-1}$)}\\
\hline
\endhead
\hline
\endfoot
\multicolumn{18}{c}{{\bf B stars}}\\
\hline
\multicolumn{9}{c|}{{\bf $\gamma$ Peg/HD 886}}                    & 32 & 2003/06/17 & 1808.384 & $-$167&213 & $-$13&2 & 4.8&0.6\\
 1 & 2002/06/21 & 1446.656 &     3&20   & $-$0.3&0.3 &	  0.6&0.1 & 33 & 2003/06/18 & 1809.369 &   131&151  & $-$21&3 &  $-$5.7&0.7\\
 2 & 2002/06/28 & 1453.648 &  $-$1&17   &    1.8&0.3 & $-$0.5&0.1 & 34 & 2003/06/19 & 1810.365 &      7&127 & $-$18&1 &  $-$1.6&0.4\\
\multicolumn{9}{c|}{{\bf $\delta$ Cet/HD 16582}}   	 	  & 35 & 2003/06/22 & 1813.367 &  $-$68&179 & $-$14&2 &     3.9&0.7\\
 1 & 2003/10/24 & 1937.501 & 40&28      &  9.9&0.2  & $-$0.8&0.1  & \multicolumn{9}{c}{{\bf $\iota$ Her/HD 160762}}		   \\
\multicolumn{9}{c|}{{\bf $\theta^2$ Ori B/HD 37042}}	          &  1 & 2001/06/18 & 1079.375 & $-$19&15   & $-$30&0.3 &   0.7&0.1\\
 1 & 2004/11/06 & 2315.694 &	143&99  &    31&1  &	0.2&0.4   &  2 & 2001/06/26 & 1087.423 &     8&14   & $-$21&0.3 &   0.5&0.1\\
\multicolumn{9}{c|}{{\bf $\eta$ Hya/HD 74280}}     	 	  & \multicolumn{9}{c}{{\bf 2 Cyg/HD 182568}}			   \\
 1 & 1998/12/17 &  164.724 &    279&116  & 20&3     &    1&1      &  1 & 2003/06/09 & 1800.442 & 1192&2225  & $-$25&20 &      4&10 \\
 2 & 1998/12/18 &  165.725 &    436&103  & 20&3	    & $-$1&1      & \multicolumn{9}{c}{{\bf BW Vul/HD 199140}}  		   \\
 3 & 1998/12/20 &  167.544 &    442&246  & 17&5	    &    3&2      &  1 & 2002/06/22 & 1447.510 &    105&347 &	 17&1  &    $-$9&1 \\
\multicolumn{9}{c|}{{\bf $\alpha$ Leo/HD 87901}}   	 	  &  2 & 2002/06/22 & 1447.526 &    536&280 &	 42&1  &   $-$13&1 \\
 1 & 1998/12/14 &  161.766 & $-$1303&848 &  21&35   &    17&16    &  3 & 2002/06/22 & 1447.543 & $-$402&260 &	 70&1  &   $-$21&1 \\
\multicolumn{9}{c|}{{\bf RS Sex/HD 89688}}	   	 	  &  4 & 2002/06/23 & 1448.525 &     92&168 &	 26&8  &   $-$18&3 \\
 1 & 1998/12/17 &  164.760 &    584&1105 &     9&12 & $-$2&5      &  5 & 2002/06/23 & 1448.541 & $-$116&148 &	 51&25 &   $-$24&9 \\
 2 & 1998/12/18 &  165.770 &   1071&1052 &    13&14 &   13&5      & \multicolumn{9}{c}{{\bf 6 Cep/HD 203467}}			   \\
\multicolumn{9}{c|}{{\bf $\alpha$ Vir/HD 116658}}  	 	  &  1 & 2002/06/19 & 1444.519 & 1518&766   & $-$10&21 &   $-$2&8  \\
 1 & 2000/07/05 &  731.353 & $-$816&408  & 81&13    & $-$46&4     & \multicolumn{9}{c}{{\bf $\pi^2$ Cyg/HD 207330}}		   \\
\multicolumn{9}{c|}{{\bf $\upsilon$ Her/HD 144206}}	          &  1 & 2001/06/22 & 1082.567 &     44&87  &  $-$8&4  &    $-$2&2 \\
 1 & 2001/07/01 & 1091.510 &	 4&18    &  4.0&0.2     & 0.1&0.1 &  2 & 2001/06/25 & 1085.567 &  $-$64&106 &  $-$8&4  &    $-$1&2 \\
 2 & 2001/07/01 & 1091.531 &	35&23    &  4.0&0.3     & 0.1&0.1 &  3 & 2001/06/25 & 1085.584 & $-$130&107 &  $-$8&4  &    $-$1&2 \\
 3 & 2001/07/01 & 1091.548 & $-$27&25    &  4.1&0.2     & 0.1&0.1 &  4 & 2001/06/27 & 1087.549 &  $-$42&139 &  $-$8&4  &    $-$1&2 \\
\multicolumn{9}{c|}{{\bf $\tau$ Her/HD 147394}} 		  &  5 & 2001/06/27 & 1087.588 &    202&156 &  $-$8&4  &    $-$1&2 \\
 1 & 2001/06/14 & 1075.399 & $-$256&160  & $-$15&1 &	 2.2&0.4  &  6 & 2001/06/29 & 1089.575 &  $-$51&134 &  $-$9&4  &    $-$1&2 \\
 2 & 2001/06/26 & 1087.371 &  $-$40&116  & $-$14&1 &	 1.6&0.4  &  7 & 2001/06/29 & 1089.592 & $-$112&135 &  $-$9&4  &    $-$1&2 \\
 3 & 2001/06/26 & 1087.394 & $-$192&127  & $-$14&1 &	 1.3&0.4  &  8 & 2001/06/30 & 1090.576 & $-$140&111 &  $-$8&4  &    $-$1&2 \\
 4 & 2002/06/12 & 1438.364 &	 80&167  & $-$18&2 &  $-$3.2&0.5  &  9 & 2001/06/30 & 1090.594 &  $-$74&126 &  $-$8&4  &    $-$1&2 \\
 5 & 2002/06/12 & 1438.380 &	506&161  & $-$19&2 &  $-$3.2&0.5  & 10 & 2002/06/13 & 1438.560 &    313&200 &  $-$9&5  &    $-$1&2 \\
 6 & 2002/06/14 & 1440.385 & $-$184&136  & $-$16&1 &	 1.7&0.4  & 11 & 2002/06/13 & 1438.576 &  $-$37&191 &  $-$9&5  &    $-$1&2 \\
 7 & 2002/06/14 & 1440.402 &	115&142  & $-$16&1 &	 1.5&0.4  & 12 & 2002/06/16 & 1441.550 & $-$116&121 &  $-$9&4  &    $-$2&2 \\
 8 & 2002/06/16 & 1442.368 & $-$128&131  & $-$16&3 &  $-$1.9&0.6  & 13 & 2002/06/16 & 1441.567 &  $-$27&113 &  $-$9&4  &    $-$2&2 \\
 9 & 2002/06/16 & 1442.384 &	 43&156  & $-$19&4 &  $-$1.0&0.9  & 14 & 2002/06/21 & 1446.501 &     51&123 &  $-$9&4  &    $-$1&2 \\
10 & 2002/06/18 & 1444.380 & $-$119& 96  & $-$15&2 &	 2.7&0.5  & 15 & 2002/06/25 & 1450.522 &  $-$36&69  &  $-$9&4  &    $-$1&2 \\
11 & 2002/06/20 & 1446.443 &	363&430  & $-$12&2 &	 1.6&0.5  & \multicolumn{9}{c}{{\bf o And/HD 217675}}			   \\
12 & 2002/06/20 & 1446.459 &   $-$5&157  & $-$13&2 &	 1.7&0.4  &  1 & 2002/06/15 & 1440.561 & 331&988    & $-$21&11  &     19&6 \\
13 & 2002/06/20 & 1446.475 &	189&140  & $-$13&2 &	 0.8&0.4  & \multicolumn{9}{c}{{\bf 1 Cas/HD 218376}}			   \\
14 & 2002/06/21 & 1447.363 &   $-$1&118  & $-$13&3 &	 2.8&0.9  &  1 & 2001/06/24 & 1084.597 & $-$111&74  &  $-$8&1  & $-$0.8&0.3\\
15 & 2002/06/21 & 1447.379 &	 97&116  & $-$13&2 &	 2.7&0.7  &  2 & 2001/06/24 & 1084.624 &  $-$72&60  &  $-$7&1  & $-$0.6&0.2\\
16 & 2002/06/23 & 1449.361 & $-$160&138  & $-$16&3 &  $-$4.7&0.5  &  3 & 2001/06/24 & 1084.641 &  $-$65&60  &  $-$7&1  & $-$0.6&0.2\\
17 & 2002/06/24 & 1450.365 &	101&108  & $-$15&1 &  $-$0.2&0.3  &  4 & 2001/07/02 & 1092.580 & $-$105&85  &  $-$8&1  & $-$0.9&0.3\\
18 & 2002/06/26 & 1452.372 & $-$266&173  & $-$13&2 &	 5.8&0.8  &  5 & 2001/07/02 & 1092.599 &     73&80  &  $-$8&1  & $-$0.9&0.3\\
19 & 2002/06/26 & 1452.392 &	 38&229  & $-$13&2 &	 6.4&0.7  &  6 & 2001/07/03 & 1093.583 &     23&60  &  $-$8&1  &    1.3&0.3\\
20 & 2003/06/07 & 1798.389 &	 24&234  & $-$18&6 &  $-$1.6&1.4  &  7 & 2001/07/03 & 1093.600 &     46&66  &  $-$9&1  &    1.3&0.3\\
21 & 2003/06/07 & 1798.406 &	242&222  & $-$16&8 &  $-$4.0&1.4  &  8 & 2002/06/11 & 1436.541 &     33&71  &  $-$6&1  & $-$0.4&0.3\\
22 & 2003/06/08 & 1799.391 &	 57&180  & $-$22&2 &  $-$7.1&0.4  &  9 & 2002/06/11 & 1436.558 &     65&69  &  $-$6&1  & $-$0.5&0.3\\
23 & 2003/06/08 & 1799.408 &	156&152  & $-$23&1 &  $-$6.3&0.4  & 10 & 2002/06/11 & 1436.574 &  $-$40&73  &  $-$6&1  & $-$0.4&0.2\\
24 & 2003/06/10 & 1801.366 & $-$141&348  & $-$12&2 &	 2.9&0.8  & 11 & 2002/06/12 & 1437.544 &  $-$13&102 &  $-$6&1  & $-$0.2&0.3\\
25 & 2003/06/10 & 1801.383 & $-$418&520  & $-$12&3 &	 2.9&0.8  & 12 & 2002/06/12 & 1437.560 &     24&92  &  $-$6&1  & $-$0.1&0.3\\
26 & 2003/06/10 & 1801.489 &	173&170  & $-$15&1 &	 2.1&0.4  & 13 & 2002/06/12 & 1437.576 & $-$165&84  &  $-$6&1  & $-$0.3&0.3\\
27 & 2003/06/11 & 1802.382 &	469&193  & $-$16&6 &	 1.9&1.4  & 14 & 2002/06/14 & 1439.571 &  $-$83&80  &  $-$6&1  & $-$0.4&0.3\\
28 & 2003/06/11 & 1802.399 &  $-$52&155  & $-$17&6 &	 4.8&2.0  & 15 & 2002/06/14 & 1439.588 &    113&90  &  $-$6&1  & $-$0.4&0.3\\
29 & 2003/06/12 & 1803.454 &  $-$12&180  & $-$17&2 &	 5.1&0.7  & 16 & 2002/06/17 & 1442.528 &    154&85  &  $-$7&1  & $-$0.4&0.3\\
30 & 2003/06/16 & 1807.373 &	390&124  & $-$14&1 &	 5.0&0.5  & 17 & 2002/06/17 & 1442.544 &     84&84  &  $-$7&1  & $-$0.4&0.3\\
31 & 2003/06/17 & 1808.366 & $-$179&209  & $-$13&2 &	 4.7&0.6  & 18 & 2002/06/17 & 1442.564 &     87&73  &  $-$7&1  & $-$0.2&0.3\\
\multicolumn{18}{c}{{\bf B supergiants}}\\
\hline
\multicolumn{9}{c|}{{\bf $\beta$ Ori/HD 34085}}                & \multicolumn{9}{c}{{\bf 67 Oph/HD 164353}}  		      \\
 1 & 2004/11/07 & 2316.509 &  $-$42&50  &    3&1  &    0.9&0.2 &  1 & 2002/06/15 & 1441.401 &  $-$33&50  & $-$3&1 &    2.3&0.4\\
 2 & 2004/11/07 & 2316.515 &   $-$4&46  &    4&1  & $-$0.1&0.2 &  2 & 2002/06/17 & 1443.417 &     49&51  & $-$1&1 &    2.6&0.4\\
 3 & 2004/11/07 & 2316.520 &   $-$7&41  &    4&1  & $-$0.1&0.2 &  3 & 2002/06/22 & 1448.411 &  $-$98&41  & $-$1&1 &    1.9&0.6\\
 4 & 2004/11/07 & 2316.531 &      1&27  &    3&1  &    0.9&0.2 &  4 & 2002/06/22 & 1448.427 &      0&40  & $-$1&1 &    1.8&0.6\\
\multicolumn{9}{c|}{{\bf $\rho$ Leo/HD 91316}}	   	       &  5 & 2002/06/26 & 1452.455 &    166&233 &     3&1 &    2.7&0.5\\
 1 & 1998/12/15 &  162.752 &    30&45    & 41&1     &  1.1&0.3 &  6 & 2002/06/26 & 1452.471 &  $-$248&85 &    3&1 &    2.6&0.4\\
 2 & 1998/12/16 &  163.736 &    11&19    & 40&1     &  0.4&0.3 & \multicolumn{9}{c}{}\\
\hline
\multicolumn{18}{c}{{\bf O stars}}\\
\hline
\multicolumn{9}{c|}{{\bf $\alpha$ Cam/HD 30614}}                  & \multicolumn{9}{c}{{\bf $\zeta$ Oph/HD 149757}}                  \\
 1 & 1998/12/14 &  161.698 &    208&177 &   19&1  & $-$3.8&0.8    &  1 & 2001/06/20 & 1081.388 & $-$1189&721  & $-$43&4 & $-$30.1&2.7 \\
 2 & 1998/12/15 &  162.672 &     37&186 & $-$1&1  & $-$3.7&0.9    &  2 & 2001/06/23 & 1083.574 & $-$1267&1692 & $-$11&4 & $-$18.0&3.3 \\
 3 & 1998/12/17 &  164.680 &      3&84  &   16&1  & $-$9.6&0.8    &  3 & 2002/06/20 & 1446.426 &    5678&3233 &  $-$7&4 &     1.6&3.7 \\
 4 & 1998/12/18 &  165.683 &    206&78  &    8&1  & $-$3.1&0.7    & \multicolumn{9}{c}{{\bf 10 Lac/HD 214680}} 		           \\
\multicolumn{9}{c|}{{\bf AE Aur/HD 34078}}                        &  1 & 1998/12/15 &  163.252 &     21&37  & $-$10&1 &    0.4&0.3 \\
 1 & 1998/12/16 &  163.680 &     54&46  &    54&1 &    0.5&0.3    &  2 & 1998/12/15 &  163.297 &     62&31  & $-$10&1 &    0.5&0.3 \\
\multicolumn{9}{c|}{{\bf $\lambda$ Ori A/HD 36861}}               &  3 & 2003/06/09 & 1799.607 &  $-$54&90  &  $-$9&1 & $-$0.4&0.3 \\
 1 & 2004/11/05 & 2315.487 & $-$47&177  &      25&1 &   1.7&1.0   &  4 & 2003/06/12 & 1802.588 &     39&69  &  $-$9&1 & $-$0.3&0.3 \\
 2 & 2004/11/06 & 2315.543 & 	45&137  &      35&1 &   1.3&0.4   &  5 & 2003/06/13 & 1803.532 &    238&90  &  $-$8&1 & $-$0.3&0.3 \\
 3 & 2004/11/06 & 2315.574 &   152&124  &      35&1 &   1.6&0.5   &  6 & 2003/06/17 & 1807.556 &     66&62  &  $-$9&1 & $-$0.2&0.3 \\
 4 & 2004/11/06 & 2315.619 & 	56&69   &      35&1 &   0.9&0.4   &  7 & 2003/06/19 & 1809.508 &     53&44  &  $-$9&1 & $-$0.3&0.3 \\
\multicolumn{9}{c|}{}                                             &  8 & 2003/06/21 & 1811.622 &      7&50  &  $-$9&1 & $-$0.4&0.3 \\
\multicolumn{9}{c|}{{\bf 15 Mon/HD 47839}}                        &  9 & 2004/11/05 & 2315.311 &  $-$45&52  &  $-$9&1 &    0.0&0.3 \\
 1 & 1998/12/14 &  161.658 &    204&215 &      36&1  & $-$0.1&0.5 & 10 & 2004/11/05 & 2315.355 &    204&55  &  $-$9&1 &    0.1&0.3 \\
 2 & 2004/11/08 & 2317.531 &  $-$44&165 &      35&1  &    0.3&0.5 & 11 & 2004/11/05 & 2315.400 &     52&57  &  $-$9&1 &    0.1&0.3 \\
 3 & 2004/11/08 & 2317.577 &  $-$92&151 &      35&1  &    0.5&0.5 & 12 & 2004/11/05 & 2315.445 &  $-$17&65  &  $-$9&1 &    0.1&0.3 \\
 4 & 2004/11/08 & 2317.636 &  $-$27&137 &      35&1  &    0.2&0.5 & 13 & 2004/11/05 & 2315.489 &     26&74  &  $-$9&1 &    0.2&0.3 \\
 5 & 2004/11/08 & 2317.705 & $-$114&180 &      34&1  &    0.7&0.5 & 14 & 2004/11/07 & 2317.396 &     49&60  &  $-$9&1 &    0.3&0.3 \\
\multicolumn{9}{c|}{}                                             & 15 & 2004/11/07 & 2317.462 &      7&52  &  $-$9&1 &    0.3&0.3 \\
\hline
\multicolumn{18}{c}{{\bf Magnetic calibrators and RR Lyrae}}\\
\hline
\multicolumn{9}{c|}{{\bf 53 Cam/HD 65339}}                        & \multicolumn{9}{c}{{\bf RR Lyrae/HD 182989}}\\
 1 & 1998/12/14 &  161.741 & $-$3607&95 & $-$3&1    &  1.7&0.4    &  1 & 2003/06/06 & 1797.483 &    128&137  & $-$54&1  & $-$0.7&0.3\\
\multicolumn{9}{c|}{{\bf $\alpha^{2}$ CVn/HD 112413}}             &  2 & 2003/06/09 & 1799.529 &    184&97   & $-$77&1  & $-$1.4&0.2\\
 1 & 2000/06/30 &  726.370 &     491&45 &    0&1    &  2.0&0.6    &  3 & 2003/06/10 & 1800.590 &    131&112  & $-$88&1  & $-$1.9&0.2\\
 2 & 2001/06/19 & 1080.362 &  $-$101&37  &    2&1   &	 0.5&0.5  &  4 & 2003/06/11 & 1801.545 &    151&178  & $-$56&1  & $-$3.7&0.3\\
 3 & 2001/06/19 & 1080.443 &   $-$14&59  &    2&1   &	 0.6&0.5  &  5 & 2003/06/13 & 1804.494 &     85&96   & $-$96&1  & $-$0.9&0.3\\
 4 & 2001/06/20 & 1081.360 &  	 598&35  &    2&1   &	 0.8&0.5  &  6 & 2003/06/16 & 1807.462 &  $-$19&80   & $-$80&1  & $-$0.6&0.2\\
 5 & 2001/07/02 & 1093.384 &  	 133&52  &    0&1   &	 1.1&0.5  &  7 & 2003/06/19 & 1810.455 &      6&68   & $-$59&1  & $-$0.2&0.2\\
 6 & 2003/06/06 & 1797.370 &  	 261&54  &    1&1   &	 1.3&0.5  &  8 & 2003/06/22 & 1812.519 &     95&104  & $-$88&1  & $-$1.7&0.3\\
\multicolumn{9}{c|}{}                                             &  9 & 2003/06/22 & 1813.456 & $-$118&104  & $-$47&1  & $-$2.0&0.3\\
\end{longtable}
}

\end{document}